\documentclass[journal]{IEEEtran}

\usepackage{cite}
\usepackage{graphicx}
\usepackage{psfrag}
\usepackage{subfigure}
\usepackage{amsmath}
\usepackage{amssymb}
\usepackage{epsfig}
\usepackage{color}
\usepackage{epsf}
\usepackage{algorithmic}
\usepackage{algorithm}
\usepackage{epsfig}

\usepackage{amsfonts}
\usepackage{times}
\usepackage{latexsym}
\usepackage{dsfont}
\usepackage{amssymb}
\usepackage{amsmath}
\usepackage{cite}
\usepackage{verbatim}
\usepackage{subfigure}

\def\bb0{{\mathbb{0}}}

\def\bb{{\mathbf{b}}}

\def\bv{{\mathbf{v}}}

\def\b0{{\mathbf{0}}}

\def\bA{{\mathbf{A}}}
\def\bB{{\mathbf{B}}}

\def\bF{{\mathbf{F}}}

\def\bH{{\mathbf{H}}}
\def\bI{{\mathbf{I}}}

\def\bR{{\mathbf{R}}}

\def\bU{{\mathbf{U}}}
\def\bV{{\mathbf{V}}}
\def\bW{{\mathbf{W}}}

\def\bbE{{\mathbb{E}}}

\def\sf0{{\mathsf{0}}}

\begin{document}

\title{The Feasibility of Interference Alignment Over Measured MIMO-OFDM Channels}
\author{Omar~El~Ayach,~\IEEEmembership{Student~Member,~IEEE,}
        Steven~Peters,~\IEEEmembership{Student~Member,~IEEE,}
        and~Robert~W.~Heath,~Jr.,~\IEEEmembership{Senior~Member,~IEEE,}
\thanks{Copyright (c) 2010 IEEE. Personal use of this material is permitted. However, permission to use this material for any other purposes must be obtained from the IEEE by sending a request to pubs-permissions@ieee.org.}
\thanks{The authors are with the Wireless Networking and Communications Group,
Department of Electrical and Computer Engineering, The University of Texas
at Austin, Austin, TX 78712 USA (e-mail: \{omarayach, speters, rheath\}@mail.utexas.edu).}
\thanks{This work was supported by the DARPA IT-MANET program, Grant W911NF-07-1-0028 and the Office of Naval Research (ONR) under grant N000141010337.}}

\markboth{IEEE TRANSACTIONS ON VEHICULAR TECHNOLOGY}%
{EL AYACH, PETERS AND HEATH: FEASIBILITY OF INTERFERENCE ALIGNMENT}

\maketitle
%


\begin{abstract}
Interference alignment (IA) has been shown to achieve the maximum achievable degrees of freedom in the interference channel. This results in sum rate scaling linearly with the number of users in the high signal-to-noise-ratio (SNR) regime. Linear scaling is achieved by precoding transmitted signals to align interference subspaces at the receivers, given channel knowledge of all transmit-receive pairs, effectively reducing the number of discernible interferers. The theory of IA was derived under assumptions about the richness of scattering in the propagation channel; practical channels do not guarantee such ideal characteristics. This paper presents the first experimental study of IA in measured multiple-input multiple-output orthogonal frequency-division multiplexing (MIMO-OFDM) interference channels. Our measurement campaign includes a variety of indoor and outdoor measurement scenarios at The University of Texas at Austin. We show that IA achieves the claimed scaling factors, or degrees of freedom, in several measured channel settings for a 3 user, 2 antennas per node setup. In addition to verifying the claimed performance, we characterize the effect of Kronecker spatial correlation on sum rate and present two other correlation measures, which we show are more tightly related to the achieved sum rate.
\end{abstract}

\section{\textbf{Introduction}} \label{sec:Intro}

\IEEEPARstart{I}{nterference} alignment (IA) is a transmission strategy for the interference channel that results in sum capacities that scale linearly, at high signal-to-noise ratio (SNR), with the number of users in the system~\cite{Jafar}. Interference alignment cooperatively aligns interfering signals over the time, space, or frequency dimensions. In multiple-input multiple-output (MIMO) interference channels, IA aligns signals in the spatial dimension by choosing transmit precoders such that interference at each receiver spans only a subspace of the receive space.  To achieve alignment and the maximum gains, however, certain dimensionality constraints need to be satisfied; alignment is only possible for a certain number of users if given a sufficient number of transmit and receive antennas. Moreover, guaranteeing the maximum degrees of freedom via precoding requires coding over infinitely many dimensions, made possible by using time or frequency extensions \cite{Jafar}, which are not considered in this paper.

MIMO interference alignment, as well as alignment in other dimensions, was first studied in ~\cite{Maddah, Jafar, JafarInt}. Since then, IA has been examined further from several angles. After deriving the high SNR sum rate scaling of IA, distributed iterative algorithms for constructing MIMO IA precoders were presented in \cite{Gomadam} and \cite{HeathIA} with varying assumptions on reciprocity and channel knowledge. Other solutions were developed for symmetric networks \cite{ConstantIA}, cellular networks \cite{Tse}, single-input single-output (SISO) networks \cite{ChoiJafChung}, and SISO networks with limited feedback \cite{LF_IA}. More recent work addresses the feasibility of IA in terms of network structure and channel state information requirements \cite{Yetis, huang-2009, TreGuiICC}. For example, \cite{Yetis} gives feasibility conditions on the number of antennas needed per node, while \cite{huang-2009} examines the possibility of applying IA to a two user network with no channel state information at the transmitter. Extending the IA concept to larger networks, \cite{johnson-2009} applies IA to large scale networks to derive new bounds on sum capacity. The work in \cite{Gomadam}-\cite{johnson-2009} has helped theoretically quantify the gains of IA, however, feasibility and performance in real channels remains an open question.

The theoretical results in \cite{Gomadam}-\cite{johnson-2009} were derived using baseband models with channels drawn independently from a continuous distribution; this represents scattering too rich to be observed in practical systems. As a result, performance may be overestimated. Moreover, there are no comprehensive interference channel measurements suitable for studying IA in practice. The only comparable results on multiuser MIMO channel measurements, not directly related to IA, target broadcast channels consisting of a single base station and several receivers, and thus do not provide the required data on measured interference channels~\cite{5.3Ghz, MU-MIMO}. Work done in \cite{SU-MU}, for example, presents multiuser measurements formed by concatenating separate single user measurements, claiming that the static measurement environment ensures the validity of the results. Related work on demonstrating IA in practice is limited to~\cite{Dina}, which tested a hybrid version of IA coupled with interference cancelation and successive decoding in a single carrier narrowband MIMO wireless local area network. The work in \cite{Dina} does not provide insight into the performance of the original MIMO IA solutions in realistic wideband channels. Moreover, \cite{Dina} downplays the importance of synchronization and other physical layer concepts in the interference channel due to its narrowband nature. Consequently, the viability of IA in measured channels has not yet been evaluated.

In this paper, we establish the feasibility of MIMO IA in slowly time varying real world channels, with no frequency or time extensions. To acquire suitable channel measurements, we implemented a MIMO-OFDM measurement testbed for the 3-user $2\times2$ MIMO interference channel, using a software defined radio platform \cite{AyaPetHea09}. We gave special attention to the proper implementation of a synchronized MIMO-OFDM physical layer, a consideration that was not emphasized in \cite{Dina, 5.3Ghz, SU-MU}, to guarantee the validity of our measurements. We augment the system we implemented in \cite{AyaPetHea09} to accommodate measurement setups over large outdoor areas. We make channel measurements for a variety of indoor and outdoor static node deployments and therefore extend the preliminary indoor results derived in \cite{AyaPetHea09}\footnote{We also obtain larger data sets, in more indoor and outdoor antenna and node configurations. The performance analysis is also extended to include various other interference channel algorithms.}. We summarize the data collected and use it to establish the true performance of IA in measured wideband channels. We examine the average sum rate achieved versus signal-to-noise ratio and show that, as predicted in theory, IA outperforms time division multiple access (TDMA) as well as other MIMO techniques. We also show that IA achieves the maximum degrees of freedom in our setup.  We characterize the effect of non-ideal propagation channel characteristics, such as Kronecker spatial correlation, on achieved sum rate. Finally, we introduce two other correlation measures, matrix collinearity and subspace distance, and show that they are more tightly related to the achieved sum rate.

In this paper we use the following notation: $\bA$ is a matrix, and $a$ is a scalar; $\bA^*$ denotes the conjugate transpose of $\bA$, $\|\bA\|_F$ is its Frobenius norm, ${\rm trace}(\bA)$ is its trace, ${\rm span}(\bA)$ is its column space, and ${\rm null} (\bA)$ its nullspace; $\nu (\bA)$ is any eigenvector of $\bA$ and $\nu_{max} (\bA)$ is the dominant eigenvector when eigenvalues are real; $\bI_N$ is the $N \times N$ identity matrix; $\mathbb{C}^N$ is the $N$-dimensional complex space.

This paper is organized as follows. Section \ref{sec:SigModel} briefly presents the MIMO-OFDM signal model in the presence of interference, Section \ref{sec:IA} summarizes the basic idea of IA and introduces several IA solutions as well as the algorithms used for comparison. Section \ref{sec:setup} details both the hardware and software used in our measurement testbed. Sections \ref{sec:results} and \ref{sec:outdoor} present and discuss the results obtained from our setup in indoor and outdoor environments respectively. We conclude with Section \ref{sec:Conclusion}.

\section{MIMO Interference Signal Model} \label{sec:SigModel}

\begin{figure} [t!]
  \centering
  \includegraphics[width=3.5in]{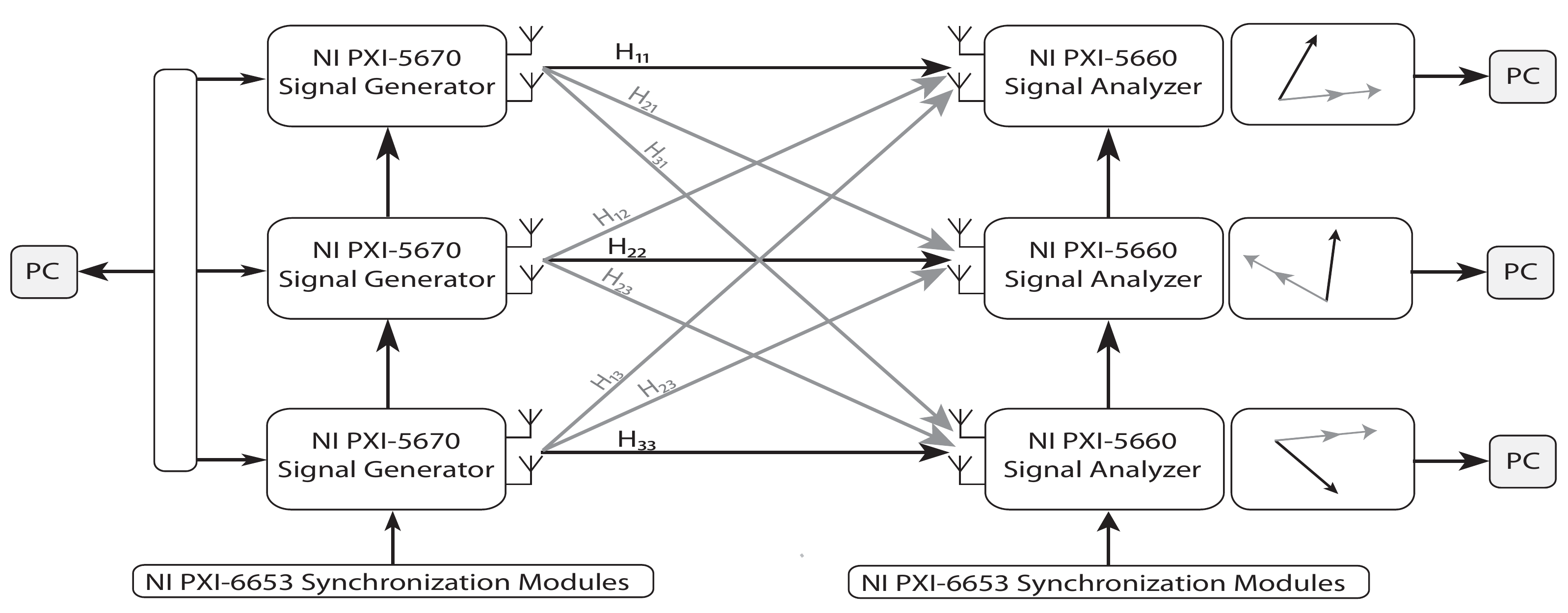}
  \caption{Simplified hardware block diagram.}
  \label{fig:blockdiag}
\end{figure}

Consider the $K$-user interference channel shown in Fig. \ref{fig:blockdiag} with $M_k$ transmit antennas at transmitter $k$ and $N_m$ receive antennas at receiver $m$. All users send $N_s$ streams of data using orthogonal frequency division multiplexing (OFDM) with $N$ subcarriers \cite{Stu-etal-2004}. This is known as MIMO-OFDM, a widely deployed transmission technique in commercial wireless systems such as IEEE 802.11n and 802.16e \cite{ZelSch04}. In the interference channel in Fig. \ref{fig:blockdiag}, each transmitter $k$ communicates with its corresponding receiver $k$ and interferes with all other receivers $m \neq k$. In this section we explain the MIMO-OFDM interference signal model in this general case, though in the remainder of this paper we specialize to the $K = 3$ user $2 \times 2$ channel in which each user sends $N_s = 1$ stream of data.

The received signal at node $k$ and subcarrier $n$ for a sufficiently slow fading channel is given by
\begin{equation}
\mathbf{y}_{k}[n] = \mathbf{H}_{k,k}[n]\mathbf{F}_{k}[n]\mathbf{s}_{k}[n] + \sum_{m \neq k} \mathbf{H}_{k,m}[n]\mathbf{F}_{m}[n]\mathbf{s}_{m}[n] + \mathbf{v}_{k}[n],
\label{eqn:intsig_model}
\end{equation}
where $\mathbf{y}_{k}$ is the $N_k \times 1$ received signal vector, $\mathbf{H}_{k,m}$ is the $N_k \times M_m$ channel matrix from transmitter $m$ to receiver $k$ with elements drawn i.i.d. from an arbitrary continuous distribution, $\mathbf{F}_{k}$ is the $M_k \times N_s$ precoding matrix used at transmitter $k$, $\mathbf{s}_{k}$ is the $N_s \times 1$ transmitted symbol vector at transmitter $k$, and $\mathbf{v}_{k}$ is a complex vector of i.i.d. circularly symmetric white Gaussian noise with covariance matrix $\bbE[\bv_k \bv_k^*]=\sigma^2\mathbf{I}_{N_k} \  \forall k$. In this signal model, we assume perfect functioning of the carrier recovery and symbol timing synchronization modules. We also assume that the impulse response of all the channels is shorter than the cyclic prefix used, thus allowing us to write the received signal as in (\ref{eqn:intsig_model}). For simplicity, the transmit power is assumed to be normalized to 1, and the effects of large scale fading are neglected.

Since the capacity region of the interference channel remains unknown, the performance of IA cannot be compared to capacity until the latter has been established. We therefore evaluate the performance of IA in comparison to other precoder designs by studying the achieved sum rate in bits/s/Hz averaged over all subcarriers with uniform power allocation \cite{BolGesPau02}. Network sum capacity is a point in the capacity region, and, more importantly, a metric that defines the total throughput of the network. The sum rate achieved by an optimal receiver, assuming ideal decoding for all precoder designs, is calculated as
\begin{eqnarray}
R_{sum}~ =~  \frac{1}{N} \sum_{n=1}^N \sum_{k=1}^{K} \log_{2} \left| \bI_{N_k} + \left(\sigma^{2}\bI_{N_k} + \bR_k[n]\right)^{-1} \right. \nonumber \\
\left. \left(\mathbf{H}_{kk}[n] \mathbf{F}_{k}[n] \mathbf{F}_{k}[n]^*\mathbf{H}_{kk}[n]^* \right)\right|,
\label{eqn:sumrateOFDM}
\end{eqnarray}
where
$$
\bR_k[n] ~ = ~ \sum_{m \neq k} \bH_{k,m}[n] \bF_m[n] \bF_m[n]^* \bH_{k,m}[n]^*
$$
is the per-subcarrier interference covariance matrix\cite{blum2003mimo}. SNR is emulated, in simulation, by varying the noise power while keeping the normalized channels constant. This normalization of the measured channels when calculating sum rate is described in Section IV-C.

\section{Interference Alignment and Other Transmit Techniques} \label{sec:IA}

In this section we summarize several transmission strategies for the interference channel. The algorithms are run offline using the measured channel data to demonstrate the expected performance in practice. We start with the closed-form solution for interference alignment, which is valid only for the three user system model when each user sends a number of streams equal to half the number of transmit and receive antennas. Then we summarize iterative interference alignment and a signal-to-interference-plus-noise-ratio (SINR) maximizing solution. We also review TDMA and greedy interference avoidance, which will be used for performance comparison. While IA is optimal in terms of sum rate scaling, and while SINR maximization outperforms IA in the low SNR regime by considering SINR in the subspace chosen for the desired signal, none of these strategies is yet proven to be sum rate optimal.

\subsection{Closed Form Interference Alignment} \label{sec:closedIA}

IA, using enough antennas per node \cite{HeathIA}, aims at choosing the set of precoding matrices $\left\{\mathbf{F}_{k}\right\}$ to force the received interference at each of the $K$ receivers to lie within a lower dimensional subspace. Specifically, if receiver $k$ intends on decoding $N_s$ independent data streams with no interference, it must restrict interference to an $N_k - N_s$ dimensional subspace of the receive signal space, $\mathbb{C}^{N_k}$.

Let $\mathbf{W}_{k}[n]$ be the $N_k \times N_s$ matrix describing the orthonormal basis for the interference free subspace used at node $k$ and subcarrier $n$. Prior to decoding, node $k$ first projects on the basis of the interference free subspace. Ignoring the AWGN noise term, this yields
\begin{multline}
\mathbf{W}_{k}[n]^*\mathbf{y}_{k}[n] = \mathbf{W}_{k}[n]^* \left( \mathbf{H}_{k,k}[n]\mathbf{F}_{k}[n]\mathbf{s}_{k}[n] +  \right. \\ 
 \left. \sum_{m \neq k} \mathbf{H}_{k,m}[n]\mathbf{F}_{m}[n] \mathbf{s}_{m}[n]\right).
\label{eqn:intalign}
\end{multline}
For alignment, the received interference must lie in the $N_k -N_s$ dimensional nullspace of $\mathbf{W}_{k}[n]^*$, which gives
\begin{equation}
\mathrm{span}(\mathbf{H}_{k,m}[n] \mathbf{F}_{m}[n]) \subseteq \mathrm{null}(\mathbf{W}_{k}[n]^*), \quad \forall m\neq k.
\label{eqn:span1}
\end{equation}
In addition to satisfying (\ref{eqn:span1}), the interference alignment solution must satisfy
\begin{equation}
\mathrm{rank}(\mathbf{W}_{k}[n]^* \mathbf{H}_{k,k}[n] \mathbf{F}_{k}[n]) = N_s
\label{eqn:intalign3}
\end{equation}
to successfully decode all $N_s$ streams with a linear receiver.

This spatial alignment approach uses a finite number of dimensions and is only proven to achieve the maximum degrees of freedom for the 3 user channel with $N_s$ equal to half the number of antennas per node \cite{Jafar}, which is the case we consider. We focus on $M_k=N_m=2,\ \forall k,m$ and $N_s=1$. In this case, the conditions for interference alignment given in (\ref{eqn:span1}) and (\ref{eqn:intalign3}) are satisfied by choosing the precoding matrices as

\begin{multline}
\mathbf{F}_{1}[n] = \nu\left((\mathbf{H}_{3,1}[n])^{-1} \mathbf{H}_{3,2}[n] (\mathbf{H}_{1,2}[n])^{-1} \right. \\
\left. \mathbf{H}_{1,3}[n] (\mathbf{H}_{2,3}[n])^{-1} \mathbf{H}_{2,1}[n]\right),
\label{eqn:f1}
\end{multline}

\begin{equation}
\mathbf{F}_{2}[n] =  (\mathbf{H}_{3,2}[n])^{-1} \mathbf{H}_{3,1}[n] \mathbf{F}_{1}[n],
\label{eqn:f2}
\end{equation}

\begin{equation}
\mathbf{F}_{3}[n] =  (\mathbf{H}_{2,3}[n])^{-1} \mathbf{H}_{2,1}[n] \mathbf{F}_{1}[n].
\label{eqn:f3}
\end{equation}

The solution presented in (\ref{eqn:f1}), (\ref{eqn:f2}), and (\ref{eqn:f3}) is not unique. In fact, any IA solution can be rotated inside its subspace without destroying alignment. Non-uniqueness can also be seen by the ability to choose any eigenvector in (\ref{eqn:f1}), each resulting in different precoders and sum rate. Since the number of IA solutions, and a method to finding the sum rate maximizing one, remains unknown, the solution space must be further investigated and non-uniqueness exploited to increase sum rate. While optimality is neither proven nor claimed, \cite{HeathIgnacio} is an example of exploiting non-uniqueness to provide a modified IA algorithm that yields better sum rate performance than in \cite{HeathIA, Jafar,Gomadam}. Finally, such closed form solutions do not yet exist for networks with more than three users, except in the case of symmetric channels and $N_s=1$ presented in \cite{ConstantIA}.


\subsection{Iterative Interference Alignment} \label{sec:iterativeIA}

In \cite{HeathIA}, alignment in a $K$-user network is formulated in a general alternating minimization framework, alternating between solving for the $K$ precoders and the $K$ interference subspaces. The alignment problem is viewed as minimizing the ``leakage'' interference power over the set of precoders $\left\{\mathbf{F}_{k}[n]\right\}$ and interference subspaces $\left\{\mathbf{C}_{k}[n]\right\}$. This minimization problem is written as
\begin{multline}
\min_{\substack{\mathbf{F}_{m}[n]^*\mathbf{F}_{m}[n]=\bI_{N_s}, \forall m \\ \mathbf{C}_{k}[n]^*\mathbf{C}_{k}[n]=\bI_{N_k-N_s}, \forall k}}  \sum_{k=1}^{K} \sum_{\substack{m=1 \\ m \neq k}}\left\| \mathbf{H}_{k,m}[n]\mathbf{F}_{m}[n]- \right. \\ \left. \mathbf{C}_{k}[n]\mathbf{C}_{k}[n]^*\mathbf{H}_{k,m}[n]\mathbf{F}_{m}[n]\right\|_{F}^{2}.
\label{eqn:valuefunc}
\end{multline}
The precoders $\left\{\mathbf{F}_{m}[n]\right\}$ are iteratively refined while keeping $\left\{\mathbf{C}_{k}[n]\right\}$ fixed, and vice versa. As a result, the pseudo code for such a minimization is
\begin{enumerate}
\item Choose the set $\left\{\mathbf{F}_{m}[n]\right\}$ randomly.
\item Choose the columns of $\mathbf{C}_{k}[n]$ to be the $N_k - N_s$ dominant eigenvectors of \\ $\sum_{m \neq k}{\mathbf{H}_{k,m}[n]\mathbf{F}_{m}[n]\mathbf{F}_{m}[n]^*\mathbf{H}_{k,m}[n]^*}$, \  $\forall k$.
\item Choose the columns of $\mathbf{F}_{m}[n]$ to be the $N_s$ least dominant eigenvectors of \\ $\sum_{k\neq m}{\mathbf{H}_{k,m}[n]^*\left(\mathbf{I}_{N_{k}} - \mathbf{C}_{k}[n]\mathbf{C}_{k}[n]^*\right)\mathbf{H}_{k,m}[n]}$, \  $\forall m$.
\item Repeat steps 2 and 3 until convergence.
\end{enumerate}
In summary, the algorithm first finds the subspaces $\mathbf{C}_{k}[n]$ which are ``closest'' to the received interference, and then calculates the precoders $\mathbf{F}_{m}[n]$ to align interference as close as possible to the found subspaces. To cancel interference using a linear receiver, for example, receiver $k$ multiplies its received signal by the orthonormal basis of $\mathbf{W}_{k}[n]=\mathbf{I}_{N_{k}} - \mathbf{C}_{k}[n]\mathbf{C}_{k}[n]^*$.

Convergence is guaranteed by the fact that steps 2 and 3 can only decrease the non-negative objective function. The non-convexity of (\ref{eqn:valuefunc}), however, implies the potential presence of multiple local optima. Thus, convergence to the global optimum is not guaranteed. To increase sum rate, this iterative algorithm, and the maximum SINR algorithm of Section \ref{sec:SINR}, can be improved by performing several random initializations and choosing the one that results in the highest sum rate \cite{PetHea:Cooperative-algorithms-for-the-MIMO-Interference:09}.

\subsection{Maximum SINR Algorithm} \label{sec:SINR}

Interference alignment does not target maximizing sum rate directly. It instead focuses on making the signal-to-interference ratio infinite at the output of the linear filters $\bW_k$; it does not attack other performance measures like the post-processing signal-to-interference-plus-noise ratio. As a result, perfect alignment often comes at the cost of lower post-alignment SNR or sum rate, the metric we are actually interested in maximizing. We can thus consider another precoder design that maximizes other metrics, perhaps without aligning interference perfectly. One such precoder design maximizes the total SINR of the network, given by
\small
\begin{multline}
S(\{\bF_k[n]\})= \\ 
\frac{\sum\limits_{k=1}^K\|\bW_k[n]\bH_{k,k}[n]\bF_k[n]\|_F^2}{\sum\limits_{k=1}^K\left(\sum\limits_{m\ne k}\|\bW_k[n]\bH_{k,m}[n]\bF_m[n]\|_F^2+\sigma^2\|\bW_k[n]\|^2 \right)},
\end{multline}
\normalsize
where $\bW_k[n]$ is now the combiner used at receiver $k$ \cite{PetHea:Cooperative-algorithms-for-the-MIMO-Interference:09}.

Instead of optimizing the sum rate or the sum SINR, which is not quite tractable, this algorithm optimizes the sum signal power over the sum interference-plus-noise power. Since the sets $\left\{\mathbf{F}_k[n]\right\}$ and $\left\{\mathbf{W}_k[n]\right\}$ are not independent, a closed-form solution for this objective function is unlikely, however, it can be solved via alternating minimization. For tractability, the precoders are constrained to have columns of equal norm, thus satisfying $\bF_k[n]^*\bF_k[n]=\frac{1}{N_s}\bI_{N_s}, \forall k$ \cite{PetHea:Cooperative-algorithms-for-the-MIMO-Interference:09}. By fixing all $\bF_m[n]$ we can solve for $\bW_k[n]$ as
\begin{multline}
\bW_k[n]=\nu_{max}\left(\left(\sum_{m\ne k}\bH_{k,m}[n]\bF_m[n]\bF_m[n]^*\bH_{k,m}[n]^* \right. \right. \\
\left. \left. + \sigma^2 \bI_{N_k}\right)^{-1}\bH_{k,k}[n] \bF_k[n]\bF_k[n]^*\bH_{k,k}[n]^*\right).
\label{eqn:maxsinrW}
\end{multline}
Conversely, by fixing all $\bW_k[n]$, we can solve for the precoders as
\small
\begin{multline}
\bF_m[n]=\nu_{max}\left(\left(\sum_{k\ne m}\bH_{k,m}[n]^*\bW_k[n]^*\bW_k[n]\bH_{k,m}[n]\right)^{-1} \right. \\ 
\left. \bH_{m,m}[n]^*\bW_m[n]^*\bW_m[n]\bH_{m,m}[n]\right).
\label{eqn:maxsinrF}
\end{multline}
\normalsize
This results in an algorithm pseudo-code given by
\begin{enumerate}
\item Choose the set $\left\{\mathbf{F}_{m}[n] \right\}$ randomly.
\item Choose the columns of $\mathbf{W}_{k}[n]$ as given by (\ref{eqn:maxsinrW}).
\item Choose the columns of $\mathbf{F}_{m}[n]$ as given by (\ref{eqn:maxsinrF}).
\item Repeat steps 2 and 3 until convergence.
\end{enumerate}
Maximum SINR can be generalized to multiple streams in bigger networks \cite{Gomadam}, where the system must be solved for each column of each matrix, resulting in non-orthogonal solutions.


\subsection{Other Transmit Strategies} \label{sec:otherBF}

For comparison, we consider other transmission schemes, such as TDMA and greedy interference avoidance.
In a network employing TDMA, transmissions from different users are orthogonal in time, meaning that only one user transmits in any given time slot. TDMA systems can take advantage of multiuser diversity by scheduling, in every time slot, the user with the most favorable channel, in terms of instantaneous rate. This requires channel information to be known at the transmitter, to make the selection process possible, and thus is a fair comparison to IA which also requires this knowledge. Note that TDMA is conceptually equivalent to other orthogonal resource allocation techniques such as FDMA, where orthogonality is in the frequency domain.

We also consider greedy interference avoidance, a beamforming strategy for the interference channel \cite{Rose_Greedy}. In SVD beamforming for the point-to-point channel \cite{PaulrajMIMO}, beamforming vectors are chosen as $\bF_k=\nu_{max}\left(\bH_{k,k}\right)$, neglecting the interference covariance matrix. In the presence of interference as in (\ref{eqn:intsig_model}), however, the rate achieved by each user depends on the matrix $\left(\sigma^{2}\bI_{N_k} + \bR_k[n]\right)^{-1/2}\left(\mathbf{H}_{k,k}[n]\right)$. In greedy interference avoidance \cite{Rose_Greedy}, users align their signals along the most dominant eigenmode of this matrix, thus sending in the direction they receive the least interference in. Since in this approach the choice of a user's precoder affects the interference subspaces observed in the network, this precoder selection is done for many iterations, in hope of reaching a fixed point, but the algorithm does not always converge \cite{YeBlum}.


\section{System Implementation and Technical Approach} \label{sec:setup}

In this section, we present the main software and hardware parts of the measurement testbed developed. We discuss the main concepts in our MIMO-OFDM system implementation such as training, channel estimation, and carrier recovery. We also introduce the system parameters used to collect channel measurements. We then discuss the main tools and metrics used in our performance analysis, as well as introduce the preliminary calculations such as the normalization needed before further processing the acquired data.

\subsection{Software Implementation}

\begin{figure} [t!]
  \centering
  \includegraphics[width=3in]{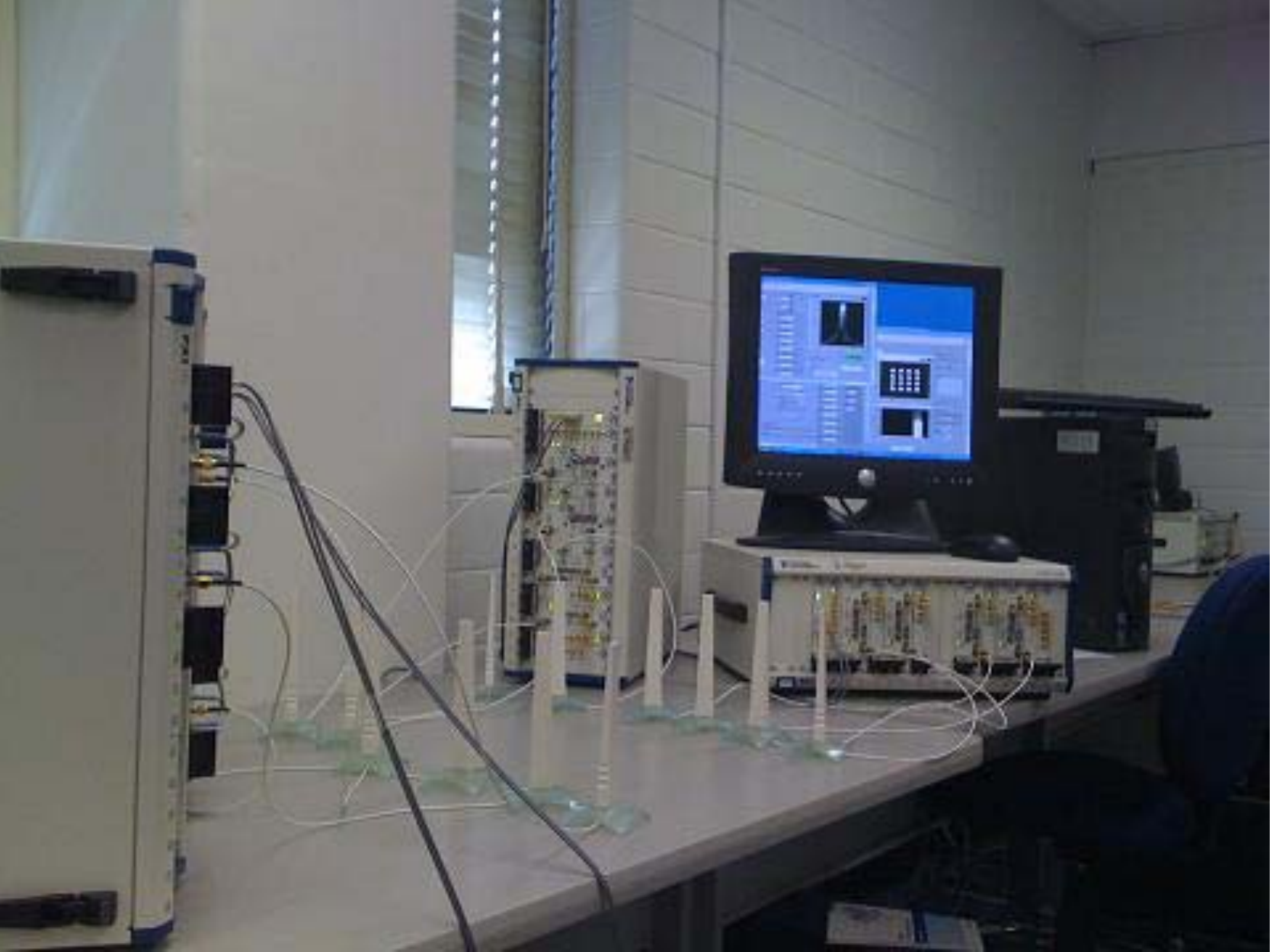}
  \caption{Picture of the measurement testbed implemented showing the antennas, RF front end equipment, as well as a sample of the software interface.}
  \label{fig:pics}
\end{figure}

Our MIMO-OFDM testbed software, implemented in National Instruments' LabVIEW \cite{LabVIEW}, uses the parameter values indicated in Table \ref{systemparam} for all communicating users in the network. We use OFDM modulation with an FFT size of 256 and a 64 sample guard interval. The total signal bandwidth used in our measurement setup is 16 MHz, which results in an effective OFDM symbol time of $20\mu s$. Communication is done at a carrier frequency of 2.4 GHz in the industrial, scientific, and medical (ISM) band. Data on each subcarrier can be modulated using BPSK or M-QAM.

\begin{figure} [t!]
  \centering
  \includegraphics[width=3.2in]{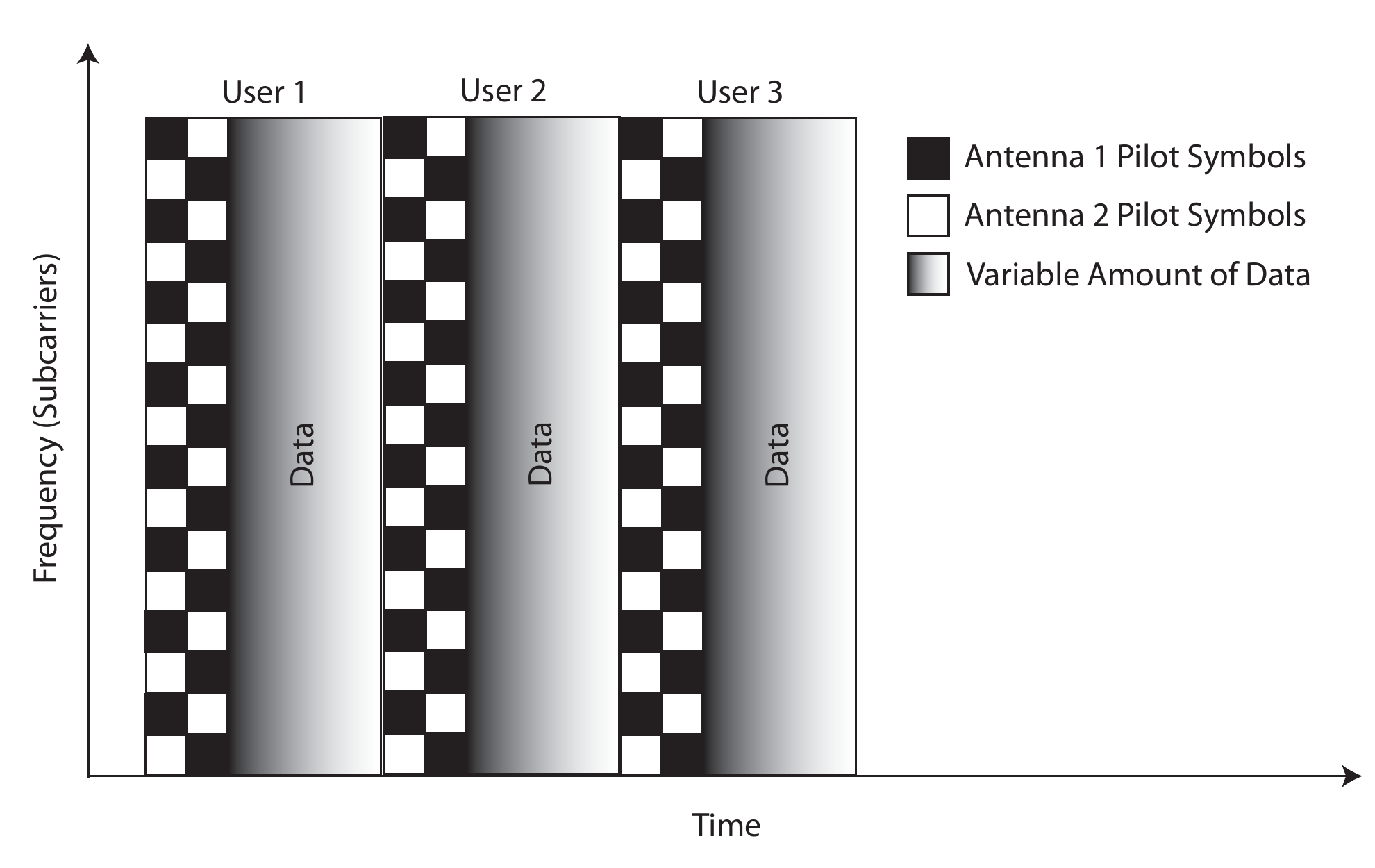}
  \caption{Simplified pilot and data example for measurement.}
  \label{fig:pilots}
\end{figure}

To faithfully predict the performance of interference alignment at high SNR, we pay special attention to our pilot structure and channel estimation implementation. Users sequentially send two OFDM symbols of frequency domain pilots \cite{Stu-etal-2004} that are known to all receivers. This makes training from each user orthogonal in time. During each user's training phase, the first symbol contains pilots from the first antenna on even subcarriers only, and the second antenna on odd ones. In the second OFDM training symbol this subcarrier assignment is reversed as shown in Fig. 3. The use of equally spaced pilots that are orthogonal across antennas is proven to be optimal in \cite{barhumi2003optimal}. We use each user's OFDM training symbols to estimate the wideband time domain channel exactly as in \cite{barhumi2003optimal} assuming proper functioning of our frequency and time synchronization modules. We also assume that the cyclic prefix length is greater than the number of channel taps in all channels. After obtaining these estimates, the optional ``payload data'' sent after each user's training is equalized using a frequency domain equalizer, similar to narrowband equalizers \cite{PaulrajMIMO}. Payload data is used to verify correct reception and estimation. For brevity, we omit the details of the channel estimation implementation in this paper and refer the reader to \cite{barhumi2003optimal} and the references therein.

We briefly discuss, however, the resulting mean square error (MSE) of our estimates, as this is a critical issue when predicting performance at SNR levels higher than our measurements' raw SNR. In a preliminary measurement campaign, we notice that our measured channels have a maximum length $L=5$ channel taps. With $M_k=2$ transmit antennas $\forall k$ and $L=5$ taps, a minimum of $LM_k=10$ pilot subcarriers are needed to estimate the channel. For the pilot structure in \cite{barhumi2003optimal}, it was shown that the resulting $MSE=\frac{\sigma^2}{P}$ where $P$ is the total power spent on training. In this case, this is the power per subcarrier multiplied by the number of pilot subcarriers. This implies that, with enough training, channel estimates can be made arbitrarily accurate. We now note that our system sends two full OFDM training symbols, instead of the needed $LM_k=10$ in practice. Therefore, after discounting null tones, we get 400 pilot subcarriers, which is forty times the needed training. This results in a channel estimate MSE which is 16dB lower than in practical systems, which often use close to minimal training. This makes the quality of our channel estimates that of a practical system functioning at a 16dB higher SNR. As a result, over training our channels by a factor of 40, allows us to faithfully predict IA performance at SNR levels 16dB higher than the measurement's raw SNR. Further details on optimal pilot structure, estimation, and mean square error can be found in \cite{barhumi2003optimal}.

Since we are mostly interested in measuring the channel, we send two pilot symbols for every OFDM payload data symbol. This puts minimum payload data in between the pilots from different users, thus keeping the measurement time in the microsecond range. Sending and equalizing payload data in the measurement exercise is recommended to verify correct reception and decoding, which ensures that the recorded channel measurements correspond to successful transmissions.

Pilot symbols are used to estimate frequency offsets between each transmit-receive pair. For proper MIMO communication, the transmit chains corresponding to the 2 transmit antennas per user are synchronized to justify the assumption of a single frequency offset per transmit-receive pair. For the sake of our channel measurements, however, and since we want to test the performance of IA in the absence of such impairments, we synchronize all users' transmit chains. Our measurements show that the use of the onboard high precision oscillators to synchronize all transmit RF chains results in frequency offsets within a 100 Hz of each other which are estimated and further corrected in software via MIMO-OFDM synchronization techniques presented in \cite{synch}. Software correction is done in stages starting with a coarse time synchronization, fractional frequency offset estimation, integral frequency offset estimation and finally fine time synchronization \cite{synch}.

\subsection{Hardware Description}

\begin{table}[t!]
\centering
\caption{MIMO-OFDM System Parameters}
\begin{tabular}{|c|c|}
\hline Carrier Freq. & 2.4 GHz \\
\hline Transmit Power & 6 dBm \\
\hline Bandwidth & 16 MHz \\
\hline FFT Size & 256 \\
\hline Subcarrier Spacing & 6.25 kHz \\
\hline Guard Interval & 64 samples \\
\hline Total Symbol Duration & 20$\mu s$ \\
\hline
\end{tabular}
\label{systemparam}
\end{table}

Our hardware setup consists of five National Instruments PXI-1045 chassis connected to 3 PCs \cite{1045}. The first PC controls 2 PXI chassis, containing the three users' transmit chains. The remaining three PXI-1045 chassis each house the receive chains of one of the users, two of which are connected to the same PC to make the testbed more mobile. In addition to the RF hardware installed, each PXI-1045 chassis holds a NI PXI-6653 module for timing and synchronization. A simplified hardware block diagram is shown in Fig. \ref{fig:blockdiag}.

Each transmitter, or RF signal generator, named PXI-5670, consists of two physical units, an arbitrary waveform generator, NI PXI-5421, and an upconverter, NI PXI-5610 \cite{5670}. The arbitrary waveform generator produces an intermediate frequency signal which is later modulated to RF via the upconverter. Each receiver, or RF signal analyzer, named NI PXI-5660, constitutes a downconverter, NI PXI-5600, and a digitizer, NI PXI-5620 \cite{5660}. On the receive side, the downconverter downconverts the signal to an intermediate frequency after which the digitizer takes over and samples the waveform which is then sent to the PC for processing using the LabVIEW software blocks. Note that each user consists of two transmit and two receive chains, totaling six transmit chains and six receive chains for our overall network setup.

Similar software defined setups have been used in papers such as \cite{RapidProto} to implement single user MIMO communication. Our system, however, is significantly more complex to support multiple users whose hardware components are housed in different chassis and controlled by different PCs. Moreover, software implementation differs greatly in the methods used for training and channel estimation as well as carrier recovery.

To support the cross chassis synchronization needed for this multi-user prototype, we install NI PXI-6653 timing and synchronization modules in each PXI-1045 chassis \cite{6653}. This module has a high stability reference oven-controlled-crystal-oscillator (OCXO) which can be exported to other chassis, thus enabling synchronization. Locking all the transmitters' phase locked loops to this high precision OCXO helps ensure minimal carrier frequency offsets between transmitters. Although frequency offset correction is implemented in software, synchronizing the transmitters in hardware further strengthens the validity of the obtained measurements. Note that due to hardware limitations, the PXI-6653 and PXI-5610 will only allow us to synchronize the intermediate frequency signal, and therefore the RF local oscillators remain independent. Our measurements indicate that the difference in frequency offsets between transmitters, when locked into the IF reference signal from the PXI-6653, is below 100 Hz at a carrier of 2.4 GHz. This remaining frequency offset is then estimated and corrected in software \cite{synch} and the MIMO links perform as expected.

In addition to synchronizing clocks, the PXI-6653 allows us to export the trigger generated when the master user begins signal generation. This digital signal is then used to trigger generation at the other transmitters. While digitally triggered acquisition, by connecting the receivers to the transmitters' PXI-6653, is possible in small-scale indoor setups such as those presented in \cite{AyaPetHea09}, our outdoor measurement setup stretches over distances of about 250ft, making digital triggering impossible. Therefore, the receivers are not connected to this reference trigger signal. To accommodate these outdoor setups, acquisition is triggered via analog edge triggers. Under this type of triggering, the receiver starts recording samples whenever the received signal level exceeds a predefined threshold. We discard any measurement that has been corrupted by the ambient interference in the 2.4 GHz ISM band and thus retain only valid interference free measurements. This is done by automatically checking the known payload data for errors and discarding any transmission with a very high bit error rate since they correspond to frames in which the synchronization and estimation blocks malfunctioned due to interference. Therefore, only measurements coming from transmissions that have been correctly received by all receivers are automatically recorded. Triggered acquisition and clock synchronization ensure that our measurements include only channel effects, and are thus free of any timing impairments.

\subsection{Technical Approach} \label{tech_approach}

We now introduce three tools that will be essential for the performance analysis that follows in Section \ref{sec:results_main}. We first discuss how channels are normalized and sum rate is calculated for our measurement scenarios. We then discuss how Kronecker spatial correlation is calculated for our measurements, a concept which we will link to the performance of IA in Section \ref{sec:results}. We then introduce two correlation metrics which we later show are more tightly related to IA and signal subspaces.

\subsubsection{Calculating Sum Rate}

Before evaluating the sum rate performance of interference alignment over measured channels, we must first obtain normalized channel matrices, $\tilde{\mathbf{H}}$. We normalize the measured channels over the full data set, i.e. no time windowing is applied. For fair comparison with the simulated Rayleigh channels, we normalize our measurements to have elements of unit variance and, thus, an average Frobenius norm of four \cite{wallace2003experimental},
\begin{equation}
\widetilde{\mathbf{H}}_{k,m}(\omega) =  2\frac{\mathbf{H}_{k,m}(\omega)}{\sqrt{\frac{1}{|\Omega|} \sum_{\omega' \in \Omega} \left|\left|\mathbf{H}_{k,m}(\omega')\right|\right|_F^2}},
\label{eqn:norm}
\end{equation}
where $\Omega$ is the set of all measurements collected in the scenario considered, i.e. when normalizing a matrix obtained when $d=1\lambda$, $\Omega$ would be the set of all channel measurements obtained in that configuration (in our measurements  $|\Omega|=50$ as indicated in Table \ref{indoor_details}).

\subsubsection{Kronecker Spatial Correlation}

The channel's spatial correlation, calculated according to the Kronecker model, is given by

\begin{equation}
\mathbf{R}_{RX} =  \frac{1}{|\Omega|}\sum_{\omega \in \Omega}\bH(\omega) \bH(\omega)^*,
\label{eqn:RXcorr}
\end{equation}

\begin{equation}
\mathbf{R}_{TX} =  \frac{1}{|\Omega|}\sum_{\omega \in \Omega}\bH(\omega)^*\bH(\omega),
\label{eqn:TXcorr}
\end{equation}
where $\Omega$ is the set of all measurements, $\mathbf{H(\omega)}$, taken in the considered configuration. When calculating correlation, the channel matrices $\mathbf{H(\omega)}$ are individually normalized to have unit Frobenius norm, $\mathbf{H(\omega)} /\left|\left|\mathbf{H(\omega)}\right|\right|_F$. For our multiuser case, when calculating receive correlation for user $k$, for example, we average over the channels $\bH_{k,\ell} \ \forall \ell$ which should have similar receive Kronecker correlation due to the separability of the model. Similarly, for the transmit correlation of user $\ell$, we consider the channels $\bH_{k,\ell} \ \forall k$. 

\subsubsection{Matrix Collinearity and Subspace Distance} 

IA performance can be closely linked to the signal spaces in the network which the Kronecker model does not fully capture. Therefore, we propose two other distance metrics that we show in Section \ref{sec:outdoor} can be used to predict performance. 

Channel matrix collinearity is a typical correlation measure considered in practice \cite{czink2100can}. The collinearity between two matrices $\bA$ and $\bB$ is defined as \cite{golub1996matrix}

\begin{equation}
c(\bA,\bB)=\frac{\left|{\rm trace}\left(\bA\bB^*\right)\right|}{\left\|\bA\right\|_F\left\|\bB\right\|_F}.
\label{eqn:collinearity}
\end{equation}
To adopt a simple correlation measure, inspired by traditional minimum distance metrics, we define maximum collinearity between cross channels as
\begin{equation}
c_{max}(\{\bH\})=\max_{(k,\ell)\neq (m,n)} c(\bH_{k,\ell},\bH_{m,n}).
\label{eqn:avg_collinearity}
\end{equation}
Considering maximum collinearity captures the worst case, most aligned, channels that negatively affect the IA solution the most.

\begin{figure} [t!]
  \centering
  \includegraphics[width=3in]{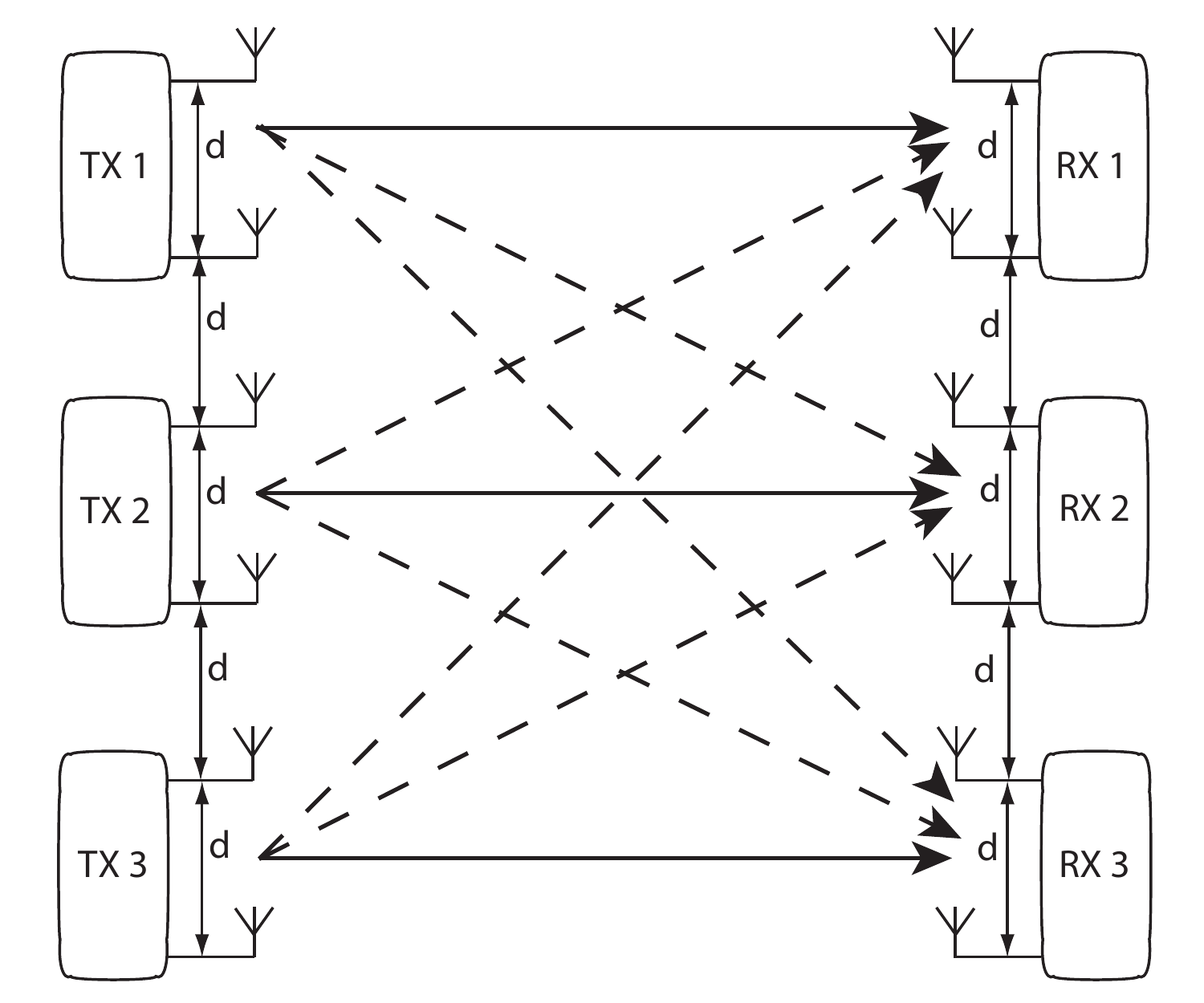}
  \caption{Schematic of an example indoor measurement configuration. Antennas are placed at varying distance $d$ apart. Also note that not only are the nodes placed at fixed positions, but all objects in this room remain in their fixed locations throughout the duration of the measurement campaign. Indoor measurement details are further summarized in Table \ref{indoor_details}.}
  \label{fig:config1}
\end{figure}

While maximum matrix collinearity is a practical correlation measure that can be linked to the performance of IA, it is sensitive to the ordering of the channels' columns. Sum rate, however, is directly linked to the SNR after projection onto the interference free space and, thus, to the distance between the subspaces spanned by the effective channels $\bH_{k,m}\bF_m$, which collinearity does not directly measure. To show this relationship in Section \ref{sec:outdoor}, we must first define the projection F-norm distance \cite{edelman} between two subspaces with orthonormal basis $\bU$ and $\bV$ as
\begin{equation}
d_{pF}\left(\bU,\bV\right)=\frac{1}{\sqrt{2}}\left|\left|\bU\bU^* - \bV\bV^*\right|\right|_F.
\label{eqn:chordal}
\end{equation}
To incorporate the distances between all channels, as well as signal and interference subspaces present in the network, we define two projection F-norm based distances, namely the average subspace distance between the set of effective channels $\bH_{k,m}\bF_m$ as
\begin{equation}
d\left(\left\{\bH\bF\right\}\right)=\sqrt{\frac{\sum\limits_{ k \neq m}d_{pF}\left(\Psi(\bH_{k,k}\bF_k),\Psi(\bH_{k,m}\bF_m)\right)^2}{{{K}\choose{K-1}}}},
\label{eqn:chordal2}
\end{equation}
and the average column space distance between the set of channels $\bH_{k,m}$
\begin{equation}
d\left(\left\{\bH\right\}\right)=\sqrt{\frac{\sum\limits_{k \neq m}d_{pF}\left(\Psi(\bH_{k,k}),\Psi(\bH_{k,m})\right)^2}{{{K}\choose{K-1}}}},
\label{eqn:chordal3}
\end{equation}
where $\Psi\left(\bA\right)$ is the operator that extracts the orthonormal basis for the column space of $\bA$. In the special case of (\ref{eqn:chordal2}), where $\bH_{k,m}\bF_m$ are vectors, $\Psi\left(\bH_{k,m}\bF_m\right)$ simply normalizes $\bH_{k,m}\bF_m$.


\section{Results} \label{sec:results_main}
In this section we present the main results on the performance of IA over measured channels. We divide the section into two subsections corresponding to our two measurement campaigns, indoor and outdoor. 

\subsection{Indoor Results} \label{sec:results}

\begin{figure} [t!]
  \centering
  \includegraphics[width=3in]{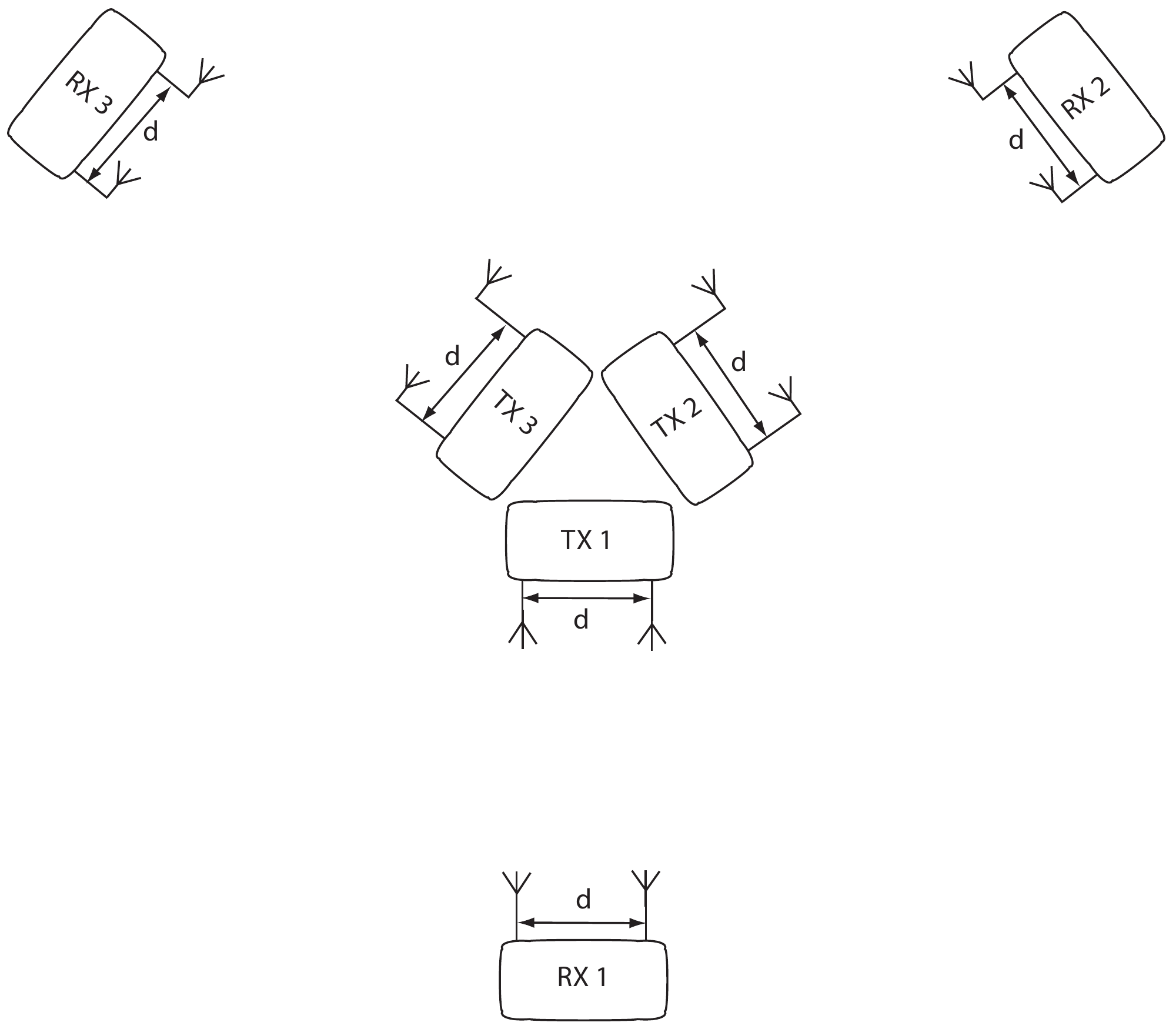}
  \caption{Schematic of an example indoor measurement configuration. Transmit antennas are all placed in the middle to model a system with co-located transmitters such as a group of access points or base stations. The receivers are then placed as shown.}
  \label{fig:config3}
\end{figure}

\begin{figure}[t!]
\centering
  \includegraphics[width=3.3in]{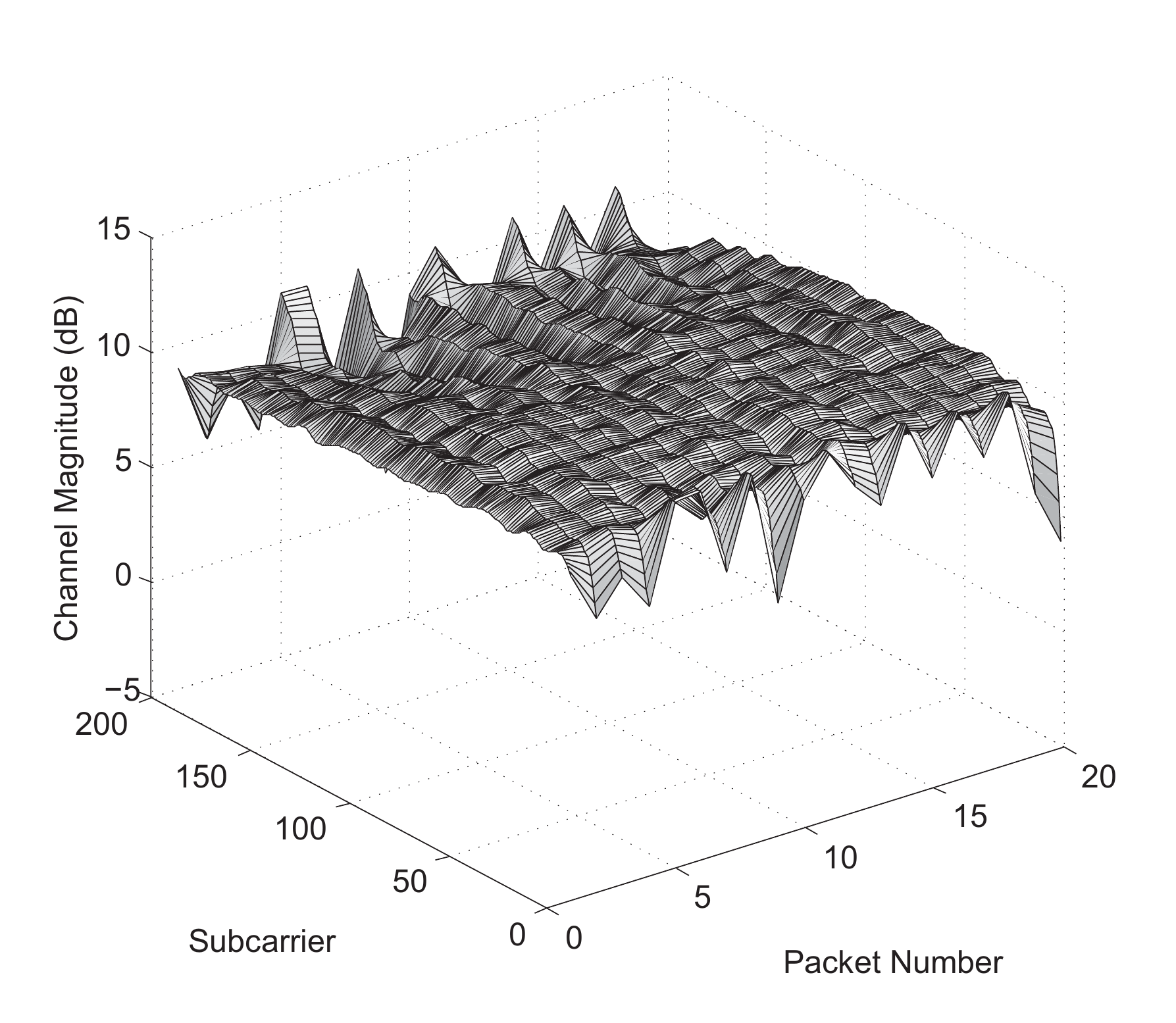}
\caption{The temporal evolution of $\|\mathbf{H_{2,1}}\|$ in our static indoor environment. $\mathbf{H_{21}}$ exhibits very limited time selectivity due to the static placement of the nodes and limited motion in the environment.}
\label{fig:timechannel}
\end{figure}

We made measurements in the Wireless Communication Lab in the Engineering Science Building at The University of Texas at Austin. Transmitter-receiver pairs were placed at distances ranging from 1 to 6 meters apart. With a wavelength of $12.5cm$, all node placements are, therefore, in the far field of the other nodes' antennas \cite{balanisantenna}. The measurement campaign details are summarized in Figs. \ref{fig:config1} and \ref{fig:config3}, and Table \ref{indoor_details}. All omnidirectional antennas are placed in the same horizontal plane with the antenna arrays placed parallel to each other as shown in the figures. SNR is kept above 25dB allowing us to predict performance up to 41dB as discussed in Section \ref{sec:setup}. Fig. \ref{fig:timechannel} shows an example temporal evolution of the channel $\|\mathbf{H}_{2,1}\|$ over 20 packet transmissions. Fig. \ref{fig:timechannel} shows two characteristics of indoor channels: limited frequency selectivity and high temporal correlation. Our measurements indicate that the channel correlation after 200 ms remains above 97\%.

Interference alignment relies on the assumption that elements of channel matrices are drawn independently at random from a continuous distribution. This assumption, however, is not likely to be satisfied in real channels that exhibit spatial correlation, thus introducing dependence in the matrix elements. While (\ref{eqn:span1}) and (\ref{eqn:intalign3}) will still be almost surely satisfiable with correlated channels, thus not influencing the feasibility of alignment, SNR after alignment may be significantly decreased due to aligned signal spaces. Our measurement results, therefore, give insight into the performance of this theoretically attractive transmit strategy in realistic channels with complexities that are not entirely captured by the simple i.i.d. models used in proving theoretical results.

\begin{figure}[t!]
  \centering
  \includegraphics[width=3.3in]{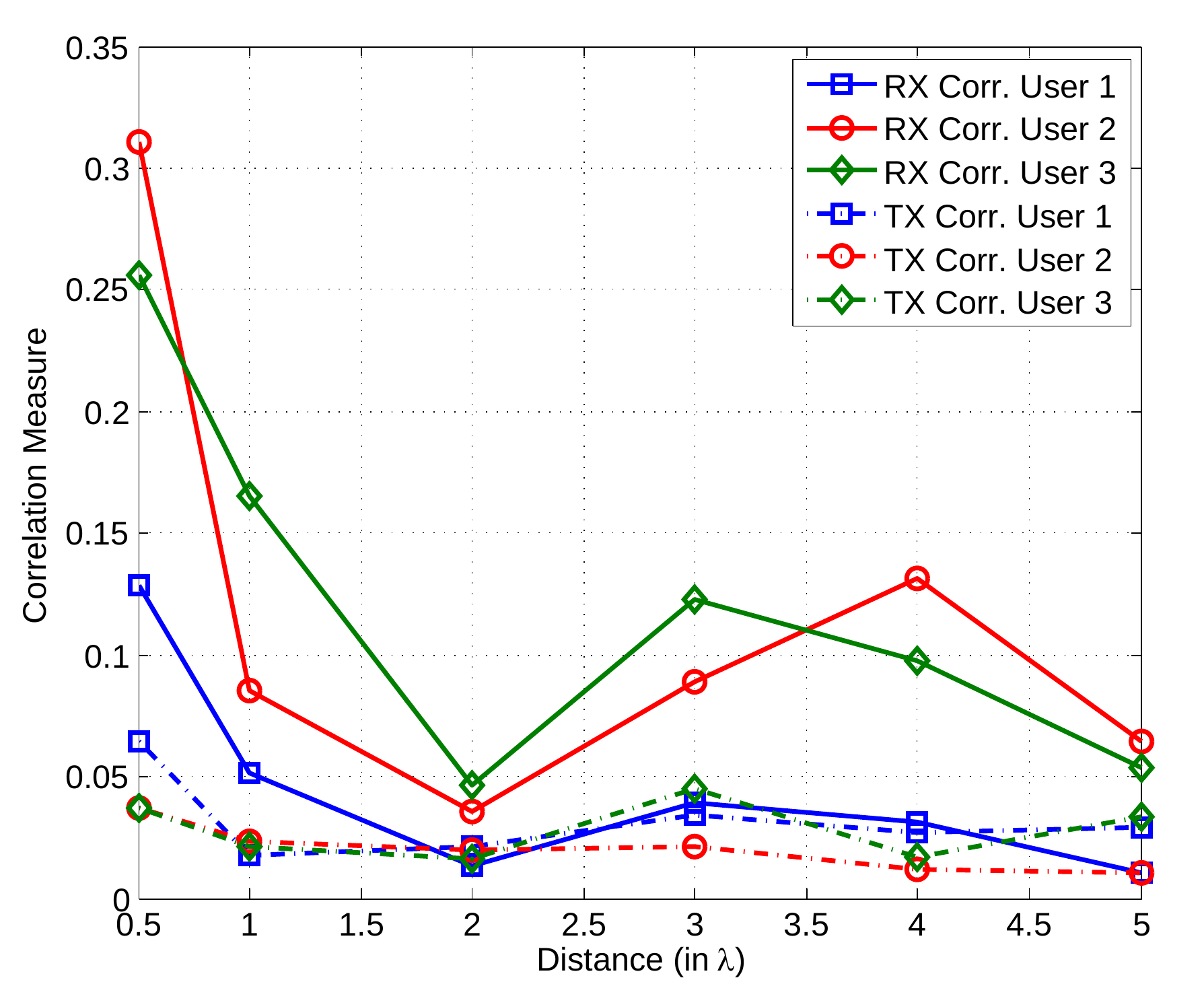}
  \caption{Channel's spatial correlation vs. antenna spacing in the indoor wireless environment.}
  \label{fig:spatial_corr}
 \end{figure}

To systematically study performance and link it to spatial correlation, we arrange nodes as shown in Fig. \ref{fig:config1} in which all antennas and users are placed a distance $d$ apart. We make measurements for variable values of $d \in \left\{\lambda/2, \lambda, 2\lambda, 3\lambda, 4\lambda, 5\lambda\right\}$ to study the effect of spatial correlation. Fig. \ref{fig:spatial_corr} shows the magnitude of the off-diagonal elements of $\bR_{RX}$ and $\bR_{TX}$ for the configuration in Fig. \ref{fig:config1}. We also conduct measurements with co-located transmitters and receivers placed on the vertices of a triangle, as shown in Fig. \ref{fig:config3}. In this arrangement, we position node pairs at a distance $1m$ apart, and antennas of the same node at a distance of $\lambda/2$, $1\lambda$, and $3\lambda$, yielding a receive correlation for user 2, for example, of 0.267, 0.117, and 0.034 respectively.

\begin{table}[t!]
\centering
\caption{Indoor Measurement Details}
\begin{tabular}{|c|c|}
\hline Tx-Rx Spacing & $\sim 6m$ \\
\hline Antenna Type & 2.4 GHz omnidirectional \\
\hline Antenna Spacing & $d \in \left\{0.5\lambda, 1\lambda, 2\lambda, \cdots, 5\lambda\right\}$ \\
\hline Configurations & 1. Equidistant nodes \& antennas \\ & 2. Triangle configuration \\
\hline \# of Measurements & 50 for each configuration \\ & \& antenna spacing \\
\hline Measurement Duration & $180\mu s$ \\
\hline Time Between Measurements & $\sim 45s$ \\
\hline Receive SNR & $> 25dB$ \\
\hline Mobility & Fixed nodes \& environment \\
\hline
\end{tabular}
\label{indoor_details}
\end{table}

\begin{figure}[t!]
\centering
  \includegraphics[width=3.3in]{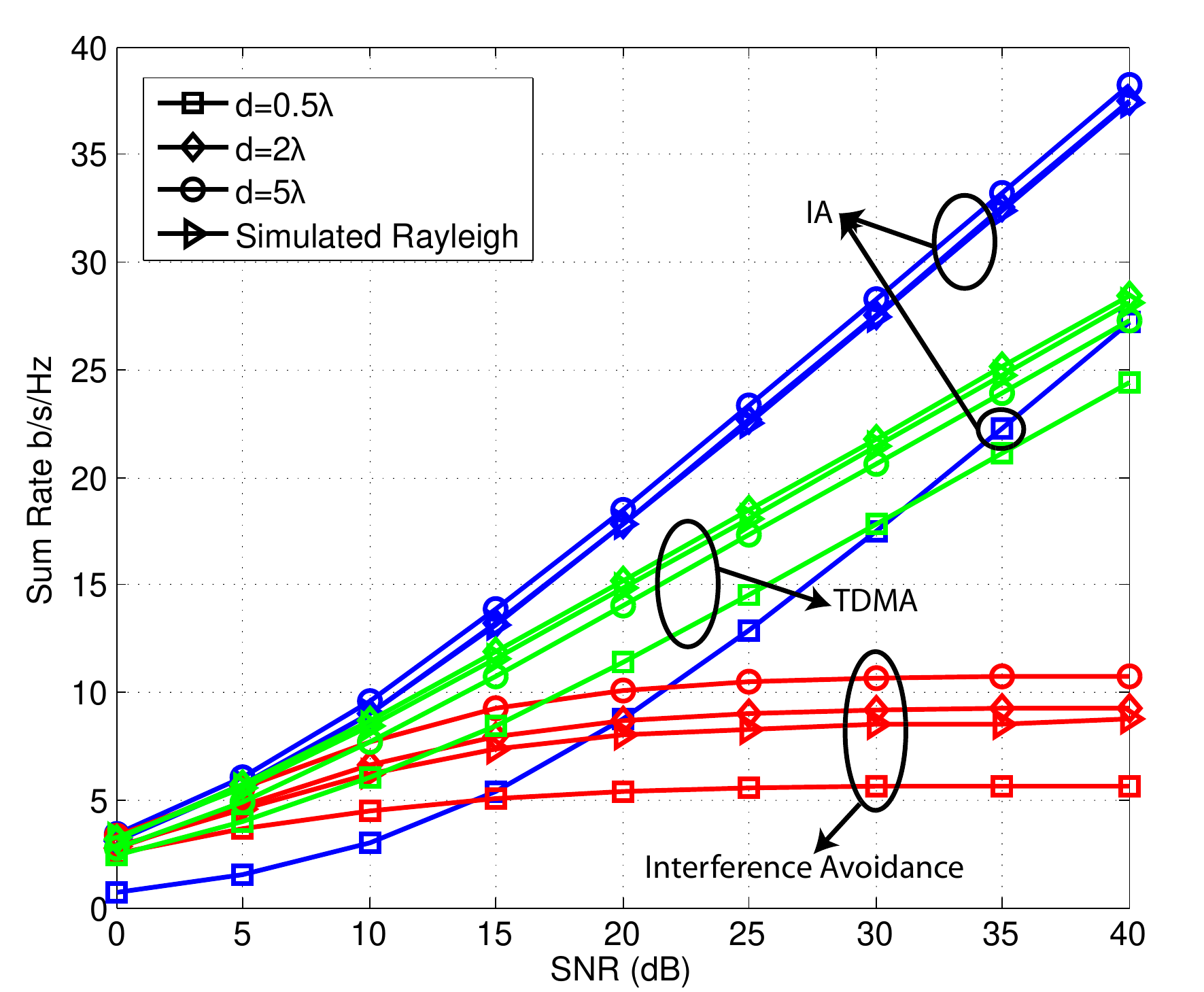}
\caption{Network sum rate vs. SNR for the configuration in Fig. \ref{fig:config1} with several antenna spacings in an indoor environment. This confirms that IA outperforms TDMA and other transmit strategies, and achieves the predicted 3 degrees of freedom for this 3 user network. We only plot a subset of scenarios since the other values of distance $d$ perform as expected and lie close to the curves of $d=2\lambda$ and $d=5\lambda$. Note: The transmission scheme for each line is annotated directly on the figure. The different measurement scenarios applied to each scheme can be identified by the markers as shown in the legend.}
\label{fig:varyall}
\end{figure}

\begin{figure}[t!]
\centering
  \includegraphics[width=3.3in]{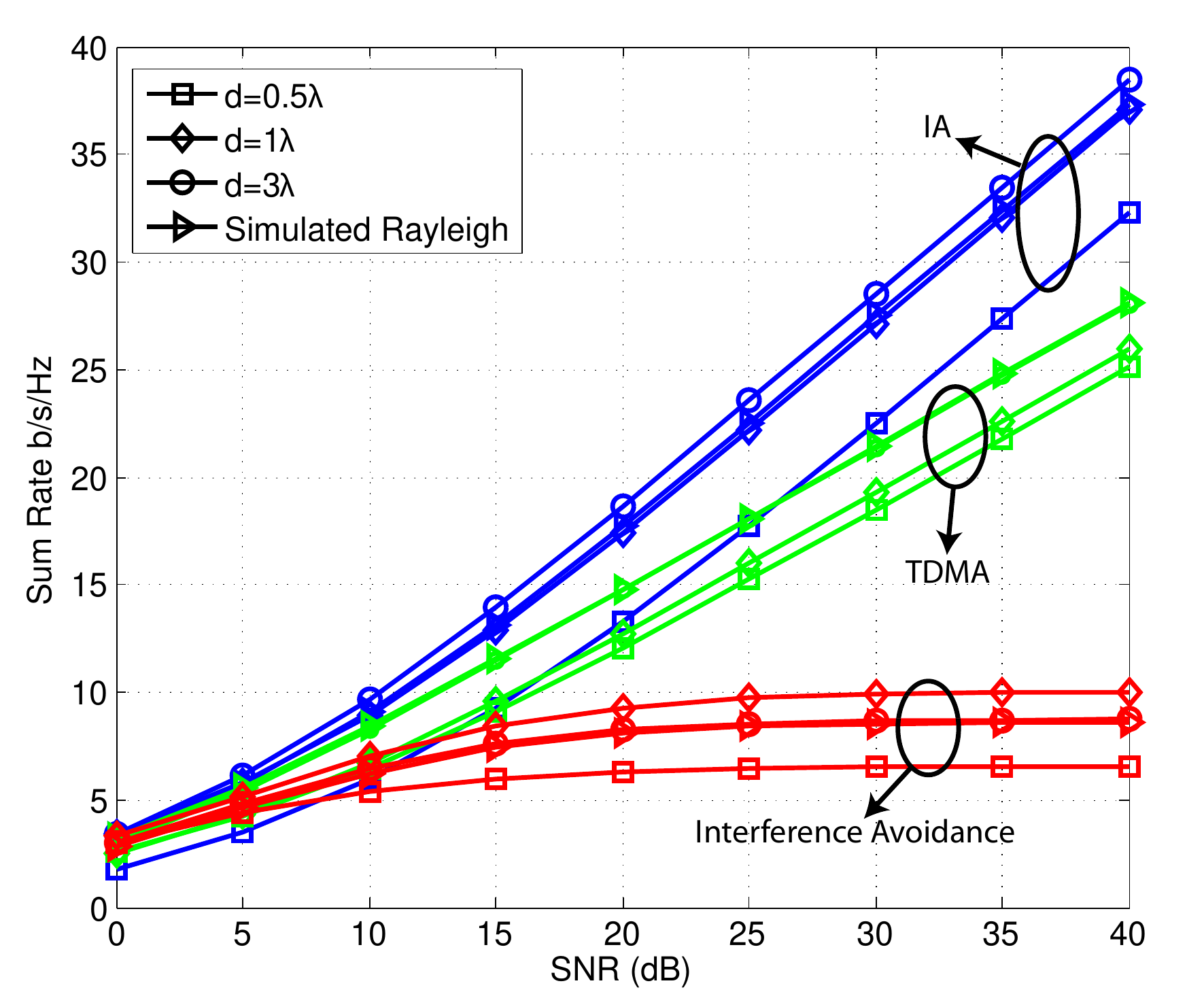}
\caption{Network sum rate for IA and TDMA Vs. SNR for the configuration with co-located transmitters shown in Fig. \ref{fig:config3} in an indoor environment. Note: The transmission scheme for each line is annotated directly on the figure. The different measurement scenarios applied to each scheme can be identified by the markers as shown in the legend.}
\label{fig:triangle}
\end{figure}

Given our measured data and the results on spatial correlation, we now turn to characterizing the performance of IA and verifying its ability to provide the maximum achievable degrees of freedom in our three user interference channel. By doing so, we will have verified the optimality of interference alignment in the high SNR regime. To that end, Fig. \ref{fig:varyall} and \ref{fig:triangle} plot the average sum rate, as defined in (\ref{eqn:sumrateOFDM}), achieved in our three user network for closed form IA and the transmit strategies presented in Section \ref{sec:otherBF}.

The indoor performance results summarized in Fig. \ref{fig:varyall}, are generated in the antenna configuration shown in Fig. \ref{fig:config1}. Fig. \ref{fig:triangle}, is generated using a configuration with co-located transmitters, as shown in Fig. \ref{fig:config3}, which may model systems having base stations or access points in a given location. As anticipated from theory, Fig. \ref{fig:varyall} and \ref{fig:triangle} verify that IA outperforms greedy interference avoidance, TDMA, and its equivalent transmission schemes. Iterative IA performs identically and, therefore, is not shown. Moreover, we note that the throughput gain from IA is largest in the high SNR regime, which is the case of claimed optimality. Comparing the rate at which network throughput increases with SNR, we observe that IA benefits more from a marginal increase of SNR, thus achieving more degrees of freedom than TDMA. The slope of the curves, relative to $\log_2 (SNR)$, is approximately 1.8 in TDMA and 2.8 for IA, thus confirming that IA provides the maximum achievable degrees of freedom, which in this case is 3.

Fig. \ref{fig:varyall} and \ref{fig:triangle} show constant differences between the various measurement scenarios at high SNR. This is due, primarily, to varying degrees of spatial correlation. Measurements with closely spaced antennas, such as the case of $d=0.5\lambda$ in both node configurations, exhibit significantly more spatial correlation across antennas, as shown in Fig. \ref{fig:spatial_corr}. This results in more aligned channels than the simulated i.i.d. Rayleigh channels, which decreases SNR after alignment. The ordering of curves in Fig. \ref{fig:varyall} reveals that IA benefits from increased antenna and user spacing with diminishing returns as users become more widely spread. This trend is consistent with the decreasing correlation shown in Fig. \ref{fig:spatial_corr}. This is also reflected in Fig. \ref{fig:triangle} which shows the performance of IA with co-located transmitters where $d=3\lambda$ outperforms $d=1\lambda$ and $d=0.5\lambda$. 

Though the trend of increasing sum rate with antenna spacing is noticeable, this is less evident at low levels of correlation. For example, in Fig. \ref{fig:varyall}, $d=5\lambda$ outperforms $d=2\lambda$, though the latter has lower Kronecker correlation. In reality, the performance of IA, is more tightly related to the distances between the signal and interference subspaces in the system. The direct link subspaces and IA performance makes antenna spacing and traditional correlation measures only a crude tool for comparison. We also note the difference in performance between the configurations of Fig. \ref{fig:config1} and \ref{fig:config3} when antenna spacing is fixed at $0.5\lambda$. As a result, the relative importance of antenna vs. user spacing requires further study. We later discuss other correlation measures that are shown, by simulation and measurement, to be more closely related to the performance of IA and can help us characterize the relative importance of both antenna and user spacing.

In addition to confirming interference alignment's theoretical achievements, our measurements give insight into the feasibility of adopting iterative algorithms in static indoor deployments. Though these algorithms may require many iterations to converge, static channels allow the precoding matrices to be used over many successive packet transmissions. This fact minimizes the relative overhead incurred by using iterative algorithms.


\subsection{Outdoor Results} \label{sec:outdoor}

\begin{figure} [t!]
  \centering
  \includegraphics[width=3.3in]{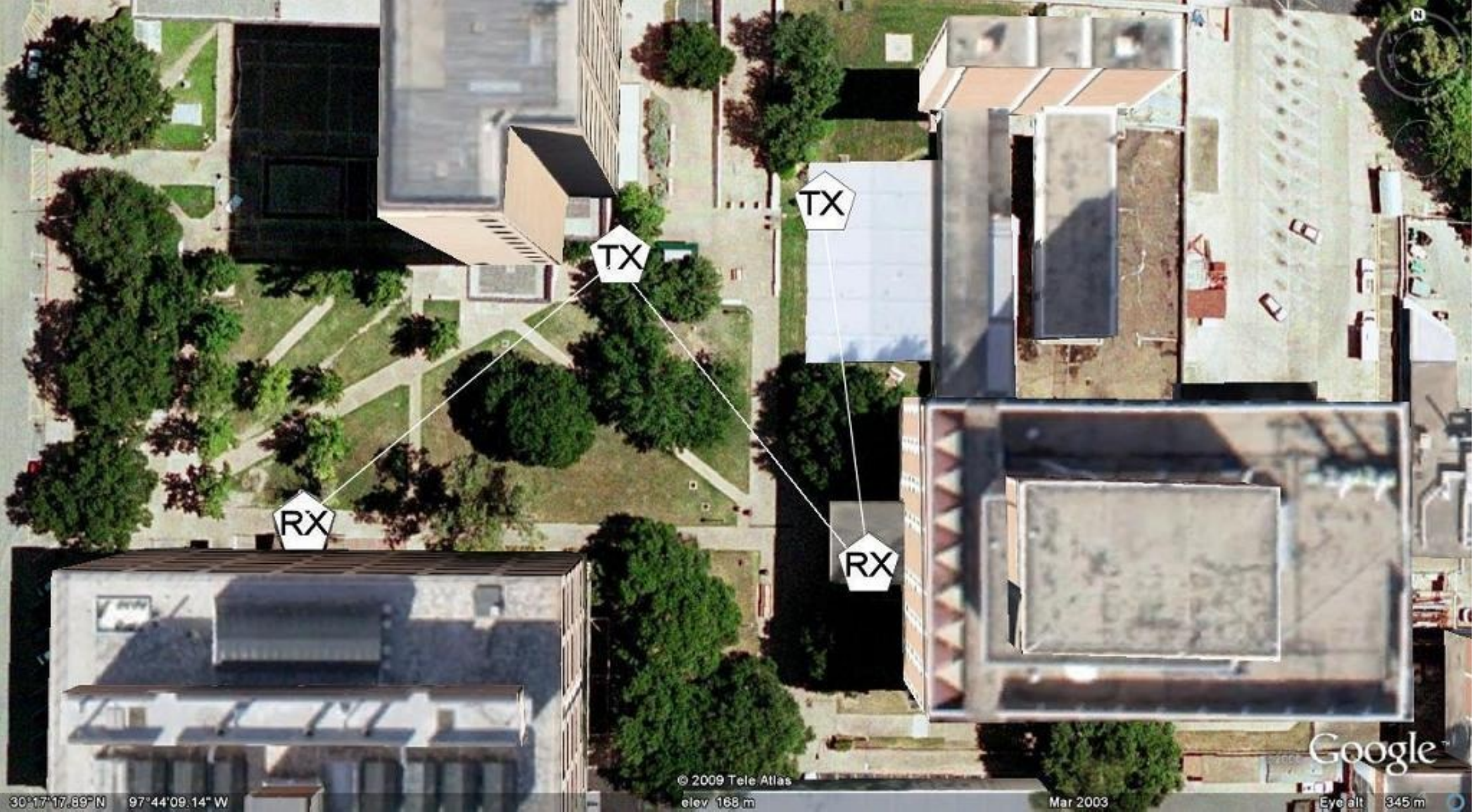}
  \caption{Area surrounding the Engineering Science Building where the outdoor measurements were taken. The TX/RX hexagons outline the approximate areas in which the 6 transceivers were placed in different configurations and do not imply actual co-location (i.e. antennas were placed around that region with sufficient antenna and user spacing).}
  \label{fig:ENSRLM}
\end{figure}

We conduct our outdoor experiments in the area surrounding the Engineering Science Building\footnote{The image is taken from Google Earth (\copyright 2009 Tele Atlas).} shown in Fig. \ref{fig:ENSRLM}. The environment contains several buildings of steel reinforced concrete, two aluminum annexes,  as well as other impeding and reflective objects normally present in a typical outdoor environment making it a good representative area to study the performance of IA in.

\begin{figure}[t!]
\centering
  \includegraphics[width=3.3in]{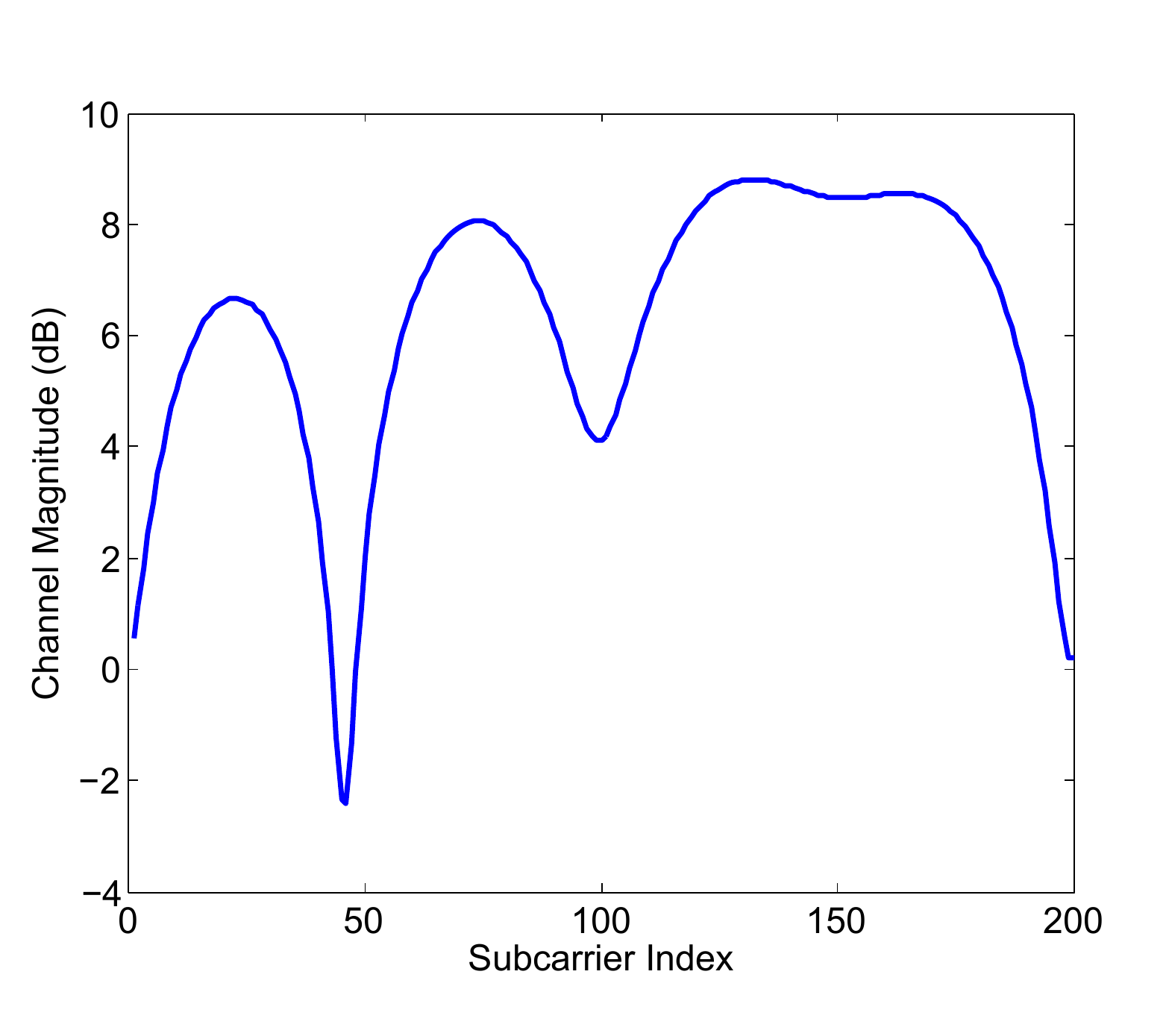}
\caption{A sample magnitude plot of an element $\bH_{1,1}$ in an outdoor NLOS arrangement. This plot shows significantly more selectivity than the channel of Fig. \ref{fig:timechannel}.}
\label{fig:outdoornlos1}
\end{figure}

Transmitters and receivers are placed approximately 200ft apart in both line-of-sight and non-line-of-sight arrangements. To support this long range transmission we use 500mW power amplifiers to maintain a receive SNR of 25dB, which allows us to predict performance up to an SNR of 41dB as shown in Section \ref{sec:setup}. Fig. \ref{fig:outdoornlos1} shows an example frequency plot of the first element of $\mathbf{H}_{1,1}$ and verifies that the outdoor channel indeed exhibits more multipath than the indoor one. Examining the power delay profile reveals the presence of 5 channel taps, resulting in a wideband channel with a coherence bandwidth of 3.2 MHz.

\begin{figure} [t!]
  \centering
  \includegraphics[width=3.3in]{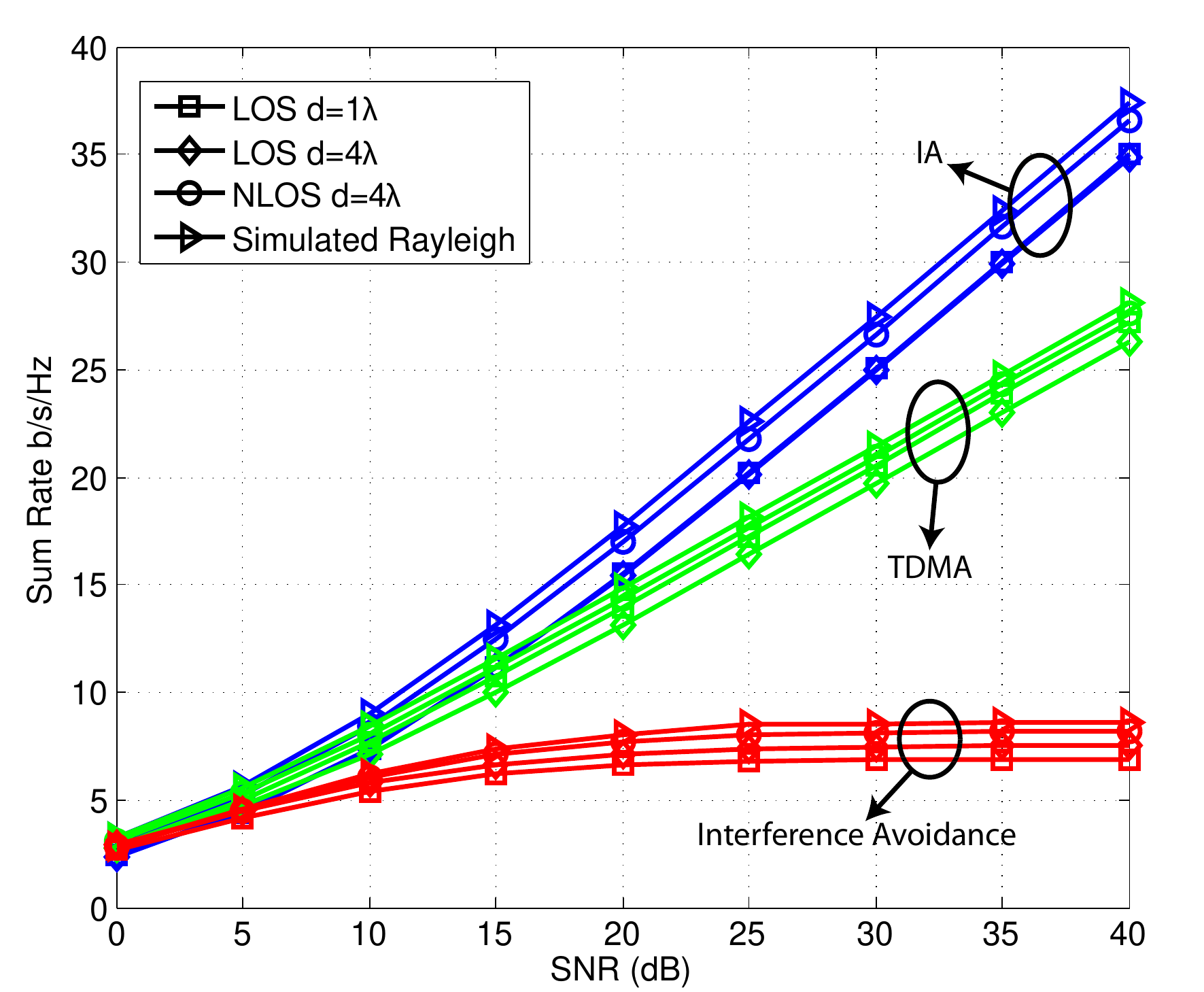}
  \caption{Network Sum Rate versus SNR for IA, TDMA, and interference avoidance in our outdoor measurement scenarios. Note: The transmission scheme for each line is annotated directly on the figure. The different measurement scenarios applied to each scheme can be identified by the markers as shown in the legend.}
  \label{fig:outdoor_IA}
\end{figure}

Fig. \ref{fig:outdoor_IA} confirms all conclusions drawn from the previous indoor results. Observing the curves in Fig. \ref{fig:outdoor_IA}, we again notice that lower correlation yields better sum rates. The NLOS scenarios, with a receive correlation coefficient for user 2, for example, of 0.105, performs significantly better than the LOS scenarios with antenna separations of $1\lambda$ and $4\lambda$, which show receive correlations of 0.166 and 0.138 respectively. Moreover, due to an increased reliance on multipath, NLOS channels vary more in space, resulting in lower correlation across users. The dependence of sum rate on correlation across users makes the Kronecker correlation coefficients not the best tool for comparison since they say very little about the distance between the signal of interest and the interference subspaces. We later show that matrix collinearity and subspace distance are more tightly related to the performance of IA and, thus, can be used to better characterize IA's performance.

Fig. \ref{fig:outdoor_IA} also confirms that IA outperforms TDMA and achieves the maximum degrees of freedom in the three user interference channel and is optimal in the high SNR regime. In addition to outperforming orthogonal techniques such as TDMA, Fig. \ref{fig:outdoor_IA} shows that when interference power is equal to the received signal power, IA outperforms greedy interference avoidance \cite{Rose_Greedy}. In many realistic ad hoc network deployments, however, communicating nodes are likely to be positioned close to one another, thus receiving different signal and interference powers. This fact makes the comparison of IA, which does not benefit from low interference power, to interference avoidance, which clearly benefits from lower interference, in a setup where all channel gains are equal, an unfair comparison. For example, one can show that at an SIR level of about 10dB, IA only outperforms interference avoidance at SNR values higher than 20dB. Therefore, depending on the received SIR levels, a network may choose to align or avoid interference. When the network is noise limited, interference is insignificant and can be avoided, but as SNR increases, communication becomes interference limited and IA dominates. IA's suboptimality at low SNR also motivates the MAX SINR algorithm described in Section \ref{sec:SINR}.

\begin{figure} [t!]
 \centering
 \includegraphics[width=3.3in]{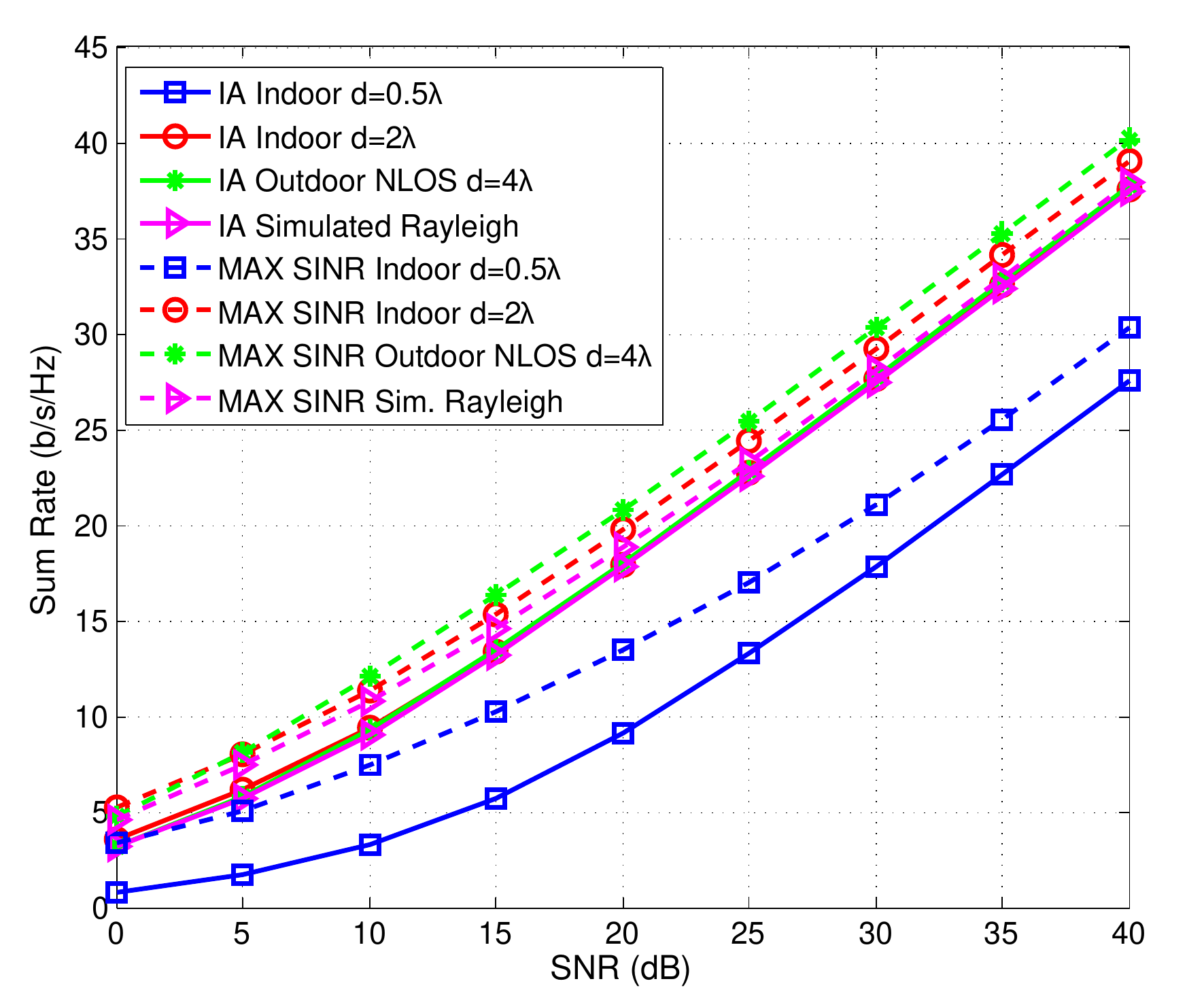}
 \caption{Sum rate Vs. SNR plots for interference alignment and the MAX SINR Algorithm. While both algorithms converge at high SNR, MAX SINR outperforms IA at low SNR when noise is a significant limiting factor in the network.}
 \label{fig:maxsinr}
\end{figure}

\begin{figure} [t!]
 \centering
 \includegraphics[width=3.3in]{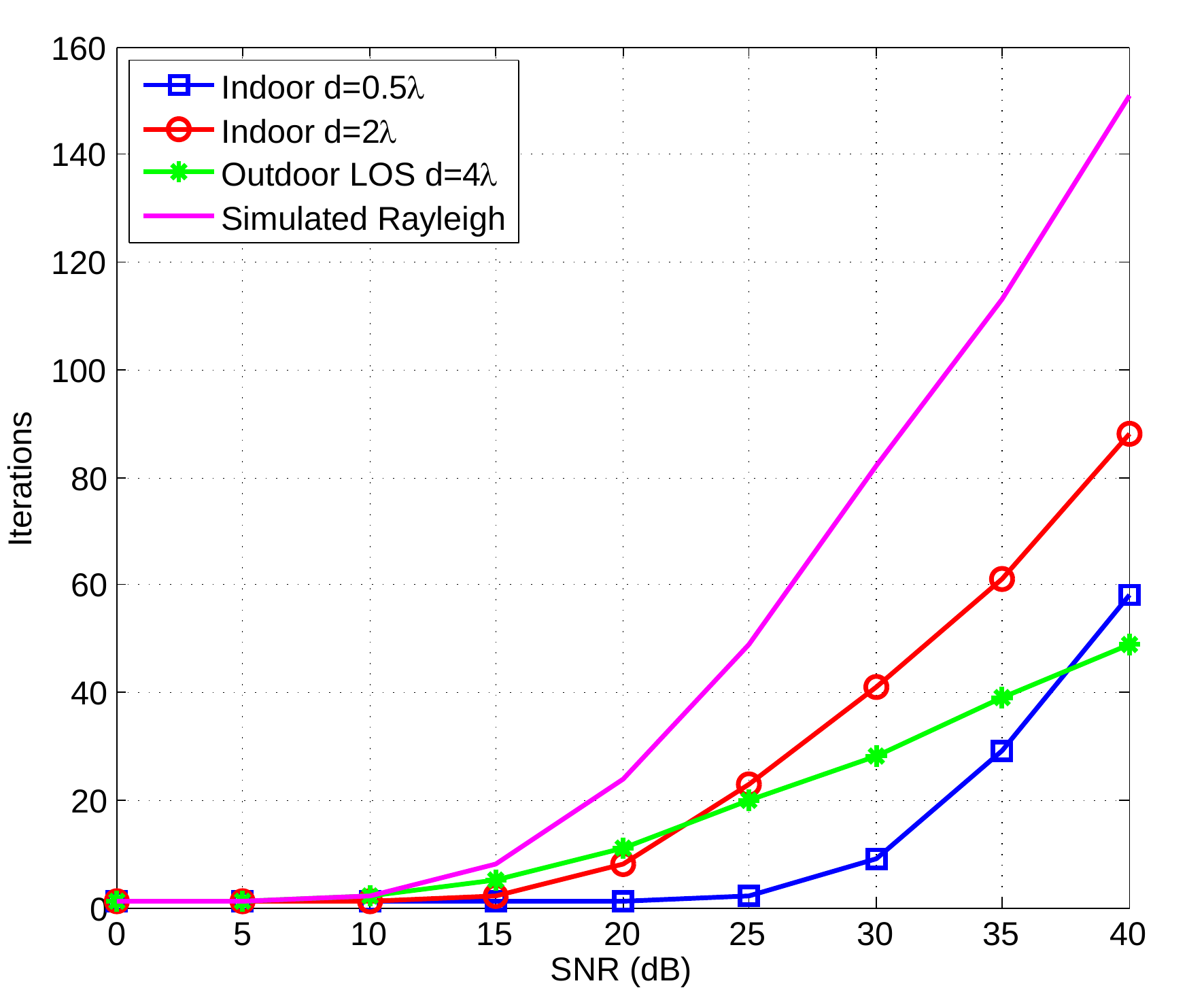}
 \caption{Number of iterations needed for MAX SINR to outperform IA. As SNR increases, the number of iterations the algorithm takes to outperforms closed form IA increases superlinearly. We note, however, that our measurements indicate that Rayleigh simulations may overestimate the number of iterations needed to outperform IA by as much as a factor 3.}
 \label{fig:intersections2}
\end{figure}

Fig. \ref{fig:maxsinr} plots the sum rates achieved by the SINR maximizing algorithm in selected indoor and outdoor scenarios. As expected, the SINR maximizing algorithm outperforms interference alignment, which is oblivious to the signal power in the interference free space. Also as expected, this performance gap is most noticeable in the low-to-medium SNR regime and decays as we transition to increasingly higher SNR. In addition to smaller improvement at high SNR, the cost of this algorithm in this regime increases as well. Fig. \ref{fig:intersections2} clearly shows the increasing number of iterations needed for MAX SINR to start outperforming closed form IA. The same can be said about absolute convergence: as SNR increases, the number of iterations needed for convergence increases. While these observations can be seen without measurements, Figs. \ref{fig:maxsinr} and \ref{fig:intersections2} show more.

From Fig. \ref{fig:maxsinr}, we see that highly correlated channels such as the case of $d=0.5\lambda$ exhibit a bigger increase in sum rate by using the MAX SINR algorithm instead of IA. This further highlights the suboptimality of IA when the i.i.d. channel assumption is farther from reality. Moreover, Fig. \ref{fig:intersections2} indicates that not only does MAX SINR outperform IA more in correlated channels, but it also does so faster. While an average of about 150 iterations are needed till MAX SINR outperforms IA in simulated Rayleigh channels at SNR=40dB, only 55 are needed when channels are highly correlated, approximately a 3 fold difference. As a result, the performance of MAX SINR is underestimated in simulation. The convergence analysis of MAX SINR and the reason why it appears to be faster and better in correlated channels is an open problem.

After presenting the immediate conclusions that can be drawn from our measurements, we return to the effect of channel correlation on sum rate. While a general trend of increasing performance with lower Kronecker correlation was observed in Fig. \ref{fig:varyall}, \ref{fig:triangle} and \ref{fig:outdoor_IA}, this is common to most MIMO techniques. IA, however, in addition to being affected by the condition of each user's channel, relies on cross channels for alignment. Therefore, correlation \emph{across} users is likely to affect IA's performance even more. We study the effect of two proposed channel ``correlation'' measures which are shown, via measurement and simulation, to more closely influence the performance of IA: channel matrix collinearity and subspace distance.

\begin{figure} [t!]
 \centering
 \includegraphics[width=3.3in]{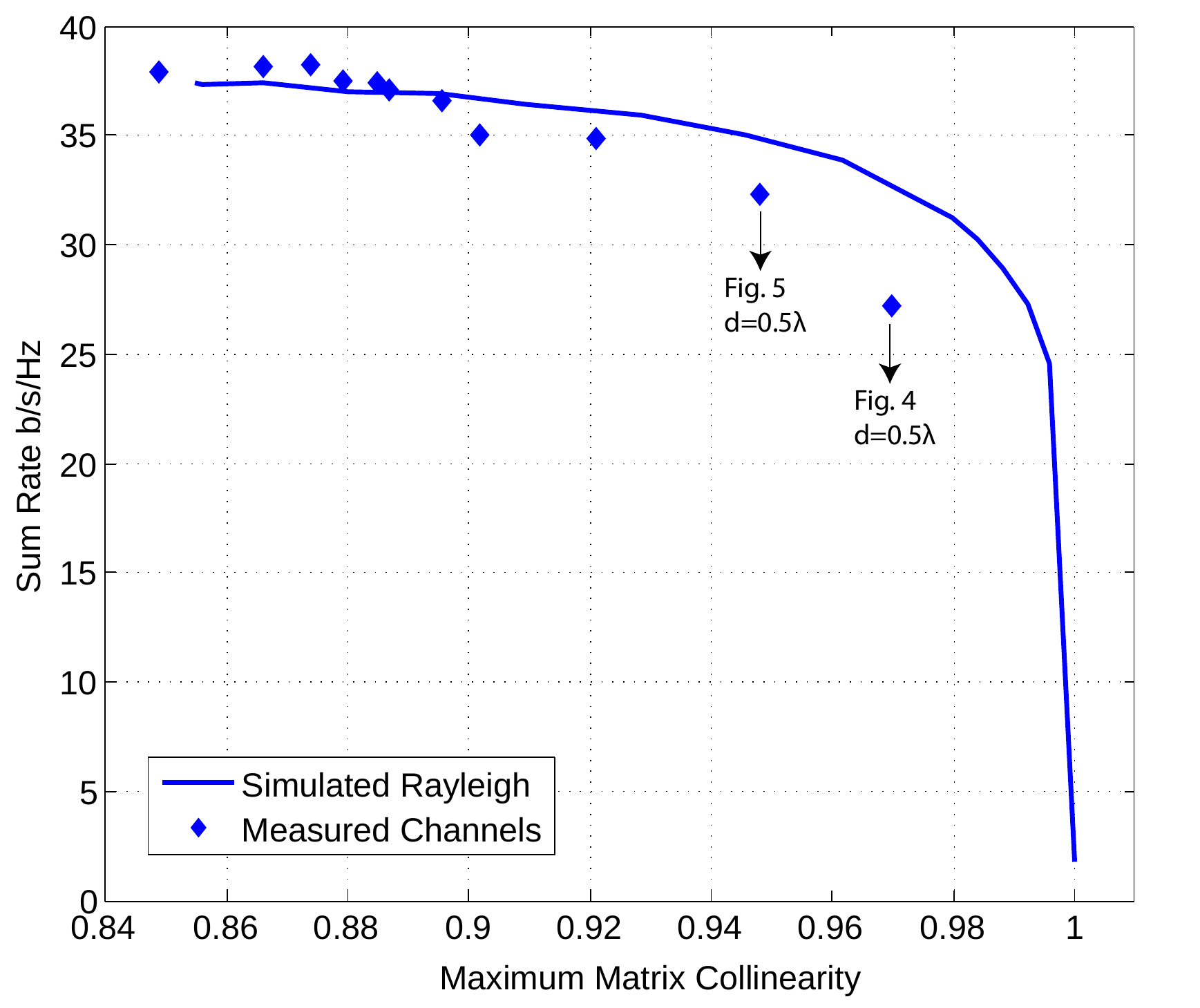}
 \caption{The effect of channel collinearity on IA performance $(SNR=40dB)$. As collinearity increases, column spaces are more aligned, which in turn reduces SNR after alignment. Our measurements indicate that as users get closer, channel collinearity increases and sum rate decreases.}
 \label{fig:collin}
\end{figure}

Fig. \ref{fig:collin} shows the performance of IA versus channel collinearity. It can be seen that all our measurements, indoor and outdoor exhibit decreasing performance with collinearity. To confirm this relationship, and since our measurements cannot span the entire range of channel collinearity, we also plot the performance over simulated Rayleigh channels with varying levels of collinearity. The match between our measurements and the simulated trend makes channel collinearity a simple feature that can be used to predict performance. Moreover, smaller values of the user spacing, $d$, result in higher collinearity and thus lower sum rate. Comparing the results for $d=0.5\lambda$, we notice that the configuration of Fig. \ref{fig:config3} outperforms that of Fig. \ref{fig:config1}. In this configuration the receivers remain separated, exhibiting lower collinearity. The relationship between sum rate and collinearity is not without reason. The factor directly controlling the achieved sum rate is SNR after alignment, which is a function of the distance between the chosen signal and interference subspaces. While collinearity is affected by the ordering of the columns of matrices, it is a measure of the similarity of column spaces. High collinearity translates into highly aligned signal subspaces and, consequently, lower SNR after alignment. For example, considering the worst case of perfectly aligned channels, we see that the precoders in (\ref{eqn:f1}), (\ref{eqn:f2}), and (\ref{eqn:f3}) yield signals that all lie in the same subspace, which drastically decreases the achieved sum rate.

\begin{figure} [t!]
 \centering
 \includegraphics[width=3.3in]{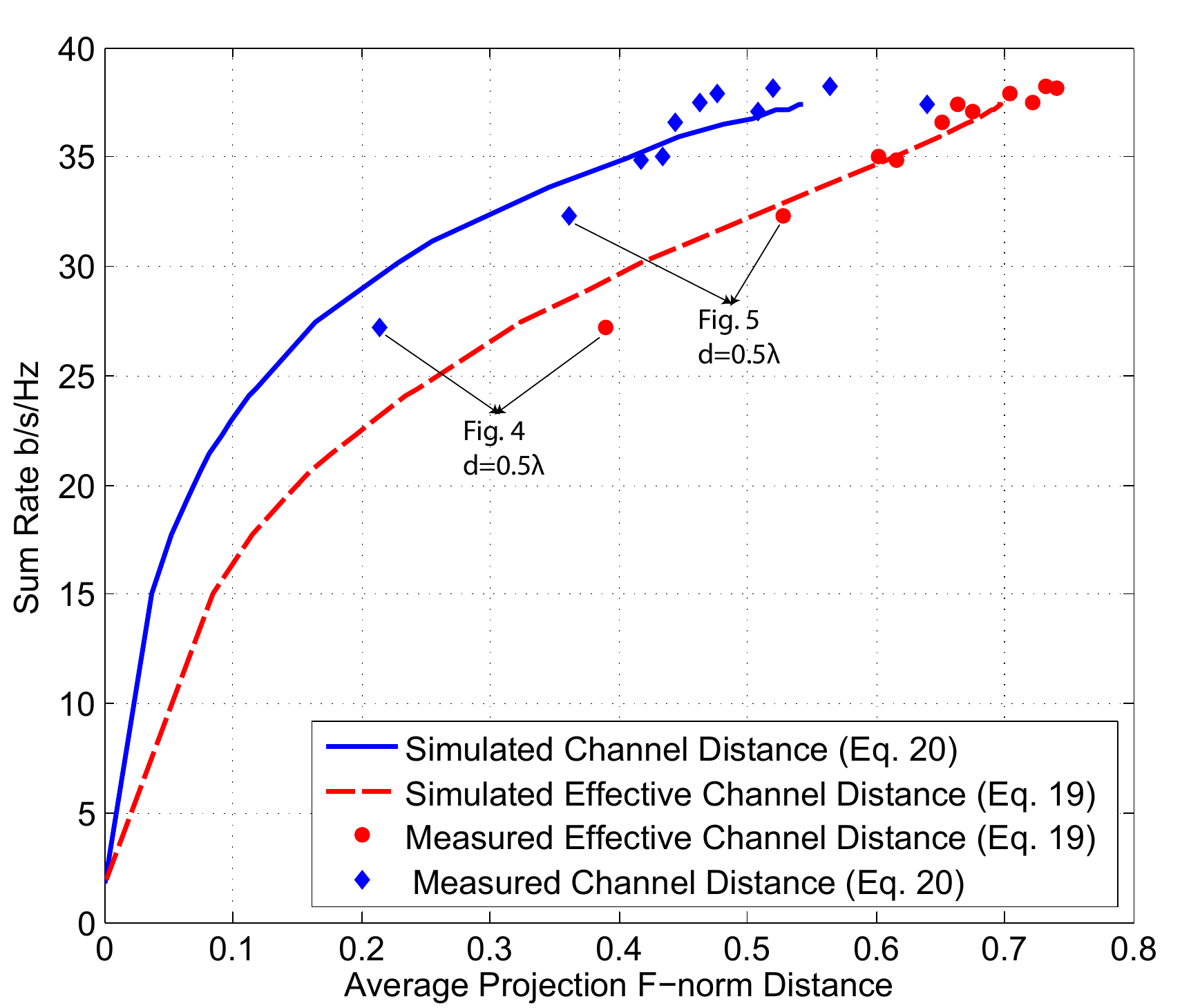}
 \caption{The effect of the average channel and effective channel distance as defined in (\ref{eqn:chordal3}) and (\ref{eqn:chordal2}) respectively, on IA performance. Our measurements indicate that as users come closer, so do the channel subspaces. As a result, both SNR after alignment and sum rate decrease.}
 \label{fig:chordal}
\end{figure}

As stated earlier, collinearity is sensitive to matrix column ordering and does not directly measure subspace distance. This motivates the use of the distances defined in (\ref{eqn:chordal2}) and (\ref{eqn:chordal3}). Fig. \ref{fig:chordal} plots the average sum rate achieved by IA over measured channels vs. the distances defined in eqs. (\ref{eqn:chordal2}) and (\ref{eqn:chordal3}). Again, to demonstrate the validity of this relationship, we plot the performance of simulated Rayleigh channels with varying subspace distance. We see that as the signal and interference subspaces become farther apart, IA's performance increases. This can be explained by considering a linear IA receiver, which projects onto the basis of the interference free space to decode. The more aligned the signal and interference spaces are, the smaller the signal component in the interference free subspace will be, diminishing post-projection SNR and sum rate.

A similar monotonic relationship between sum rate and the distance between the subspaces of the channels themselves, defined in (\ref{eqn:chordal3}), confirms the intuition that closer aligned channels result in closer aligned signal and interference subspaces. The metrics used in Fig. \ref{fig:chordal}, though more complicated, are better estimators of the performance than matrix collinearity. This can be seen by noting that the curves in Fig. \ref{fig:chordal}, have a well behaved derivative over a larger subset of their domain, i.e. have a less pronounced cut off behavior for highly correlated channels. As a result, slight changes in subspace distance, even close to zero, can more accurately estimate incremental changes in sum rate. We end by saying that in summary we have showed that ``cross-user correlation'' influences sum rate more directly. Therefore, IA is expected to perform better when minimizing cross-correlation between channels takes priority over the typical antenna separation. Although the connection between user spacing and collinearity or subspace distance is not in the scope of this paper, our results indicate that users positioned close to each other, in general, result in worse channels, from an IA perspective.


\section{Conclusion and Future Work} \label{sec:Conclusion}

In this paper, we have presented the first MIMO interference channel testbed programmed using a flexible software defined radio base. We have presented indoor and outdoor network channel measurements collected in our university environment, and then processed them to evaluate the performance gains of IA. We showed that the observed gains are in agreement with those found through theory and simulation. We also characterized the effect of channel imperfections on the sum rate achieved by IA. In subsequent work we will extend the measurement setup beyond three users as well as implement a real time closed loop IA system. This effort will be a major step in transforming IA from a theoretical concept, to a solution for large scale ad hoc networks.


\bibliographystyle{IEEEtran}
\bibliography{IEEEabrv,IA_TransVT}

\begin{thebibliography}{10}
\providecommand{\url}[1]{#1}
\csname url@samestyle\endcsname
\providecommand{\newblock}{\relax}
\providecommand{\bibinfo}[2]{#2}
\providecommand{\BIBentrySTDinterwordspacing}{\spaceskip=0pt\relax}
\providecommand{\BIBentryALTinterwordstretchfactor}{4}
\providecommand{\BIBentryALTinterwordspacing}{\spaceskip=\fontdimen2\font plus
\BIBentryALTinterwordstretchfactor\fontdimen3\font minus
  \fontdimen4\font\relax}
\providecommand{\BIBforeignlanguage}[2]{{%
\expandafter\ifx\csname l@#1\endcsname\relax
\typeout{** WARNING: IEEEtran.bst: No hyphenation pattern has been}%
\typeout{** loaded for the language `#1'. Using the pattern for}%
\typeout{** the default language instead.}%
\else
\language=\csname l@#1\endcsname
\fi
#2}}
\providecommand{\BIBdecl}{\relax}
\BIBdecl

\bibitem{Jafar}
V.~Cadambe and S.~Jafar, ``Interference alignment and degrees of freedom of the
  {K}-user interference channel,'' \emph{IEEE Trans. Inf. Theory}, vol.~54,
  no.~8, pp. 3425--3441, August 2008.

\bibitem{Maddah}
M.~Maddah-Ali, A.~Motahari, and A.~Khandani, ``Signaling over {MIMO} multi-base
  systems: combination of multi-access and broadcast schemes,'' \emph{Proc. of
  IEEE International Symposium on Information Theory, Seattle, WA}, pp.
  2104--2108, July 2006.

\bibitem{JafarInt}
S.~Jafar and M.~Fakhereddin, ``Degrees of freedom for the {MIMO} interference
  channel,'' \emph{Proc. of IEEE International Symposium on Information Theory,
  Seattle, WA}, pp. 1452--1456, July 2006.

\bibitem{Gomadam}
K.~Gomadam, V.~Cadambe, and S.~Jafar, ``Approaching the capacity of wireless
  networks through distributed interference alignment,'' \emph{Proc. of IEEE
  Global Telecommunications Conference, New Orleans, LA}, pp. 1--6, December
  2008.

\bibitem{HeathIA}
S.~W. Peters and R.~W. Heath, Jr., ``Interference alignment via alternating
  minimization,'' \emph{Proc. of IEEE International Conference on Acoustics,
  Speech, and Signal Processing}, pp. 2445 --2448, April 2009.

\bibitem{ConstantIA}
R.~Tresch, M.~Guillaud, and E.~Riegler, ``On the achievability of interference
  alignment in the {K-User} constant {MIMO} interference channel,'' \emph{Proc.
  of IEEE Workshop on Statistical Signal Processing}, pp. 277--280, September
  2009.

\bibitem{Tse}
C.~Suh and D.~Tse, ``Interference alignment for cellular networks,''
  \emph{Proc. of Allerton Conference on Communication, Control, and Computing,
  Monticello, IL}, Sept. 2008.

\bibitem{ChoiJafChung}
S.~W. Choi, S.~Jafar, and S.-Y. Chung, ``On the beamforming design for
  efficient interference alignment,'' \emph{Communications Letters, IEEE},
  vol.~13, no.~11, pp. 847 --849, nov. 2009.

\bibitem{LF_IA}
I.~Thukral and H.~Bolcskei, ``Interference alignment with limited feedback,''
  \emph{Proc. of IEEE International Symposium on Information Theory, Seoul,
  Korea}, July 2009.

\bibitem{Yetis}
C.~Yetis, T.~Gou, S.~Jafar, and A.~Kayran, ``Feasibility conditions for
  interference alignment,'' \emph{Global Telecommunications Conference, 2009.
  GLOBECOM 2009. IEEE}, pp. 1 --6, nov. 2009.

\bibitem{huang-2009}
\BIBentryALTinterwordspacing
C.~Huang, S.~A. Jafar, S.~Shamai, and S.~Vishwanath, ``On degrees of freedom
  region of {MIMO} networks without {CSIT},'' 2009. [Online]. Available:
  \url{http://www.citebase.org/abstract?id=oai:arXiv.org:0909.4017}
\BIBentrySTDinterwordspacing

\bibitem{TreGuiICC}
R.~Tresch and M.~Guillaud, ``Cellular interference alignment with imperfect
  channel knowledge,'' \emph{IEEE International Conference on Communications
  (ICC), Workshop on LTE Evolution, Dresden, Germany}, June 2009.

\bibitem{johnson-2009}
M.~Aldridge, O.~Johnson, and R.~Piechocki, ``Asymptotic sum-capacity of random
  gaussian interference networks using interference alignment,''
  \emph{Information Theory Proceedings (ISIT), 2010 IEEE International
  Symposium on}, pp. 410 --414, jun. 2010.

\bibitem{5.3Ghz}
J.~Koivunen, P.~Almers, V.-M. Kolmonen, J.~Salmi, A.~Richter, F.~Tufvesson,
  P.~Suvikunnas, A.~Molisch, and P.~Vainikainen, ``Dynamic multi-link indoor
  {MIMO} measurements at 5.3{GHz},'' \emph{Proc. of European Conference on
  Antennas and Propagation, Edinburgh, UK}, pp. 1--6, November 2007.

\bibitem{MU-MIMO}
F.~Kaltenberger, M.~Kountouris, D.~Gesbert, and R.~Knopp, ``On the tradeoff
  between feedback and capacity in measured {MU-MIMO} channels,'' \emph{IEEE
  Trans. Wireless Commun.}, vol.~8, no.~9, pp. 4866--4875, September 2009.

\bibitem{SU-MU}
G.~Bauch, J.~Bach~Andersen, C.~Guthy, M.~Herdin, J.~Nielsen, J.~Nossek,
  P.~Tejera, and W.~Utschick, ``Multiuser {MIMO} channel measurements and
  performance in a large office environment,'' \emph{Proc. of IEEE Wireless
  Communications and Networking Conference, Hong Kong}, pp. 1900--1905, March
  2007.

\bibitem{Dina}
S.~Gollakota, S.~D. Perli, and D.~Katabi, ``Interference alignment and
  cancellation,'' \emph{SIGCOMM Computer Communication Review}, vol.~39, no.~4,
  pp. 159--170, 2009.

\bibitem{AyaPetHea09}
O.~El~Ayach, S.~W. Peters, and R.~W. Heath, Jr., ``Real world feasibility of
  interference alignment using {MIMO-OFDM} channel measurements,'' \emph{Proc.
  of IEEE Conference on Military Communications, Boston, MA}, pp. 1--6, October
  2009.

\bibitem{Stu-etal-2004}
G.~Stuber, J.~Barry, S.~McLaughlin, Y.~Li, M.~Ingram, and T.~Pratt, ``Broadband
  {MIMO-OFDM} wireless communications,'' \emph{Proceedings of the IEEE},
  vol.~92, no.~2, pp. 271--294, Feb 2004.

\bibitem{ZelSch04}
A.~van Zelst and T.~Schenk, ``Implementation of a {MIMO OFDM-based} wireless
  lan system,'' \emph{IEEE Trans. Signal Process.}, vol.~52, no.~2, pp.
  483--494, Feb. 2004.

\bibitem{BolGesPau02}
H.~Bolcskei, D.~Gesbert, and A.~Paulraj, ``On the capacity of wireless systems
  employing {OFDM}-based spatial multiplexing,'' \emph{IEEE Trans. Commun.},
  vol.~50, pp. 225--234, February 2002.

\bibitem{blum2003mimo}
R.~Blum, ``{MIMO capacity with interference},'' \emph{IEEE Journal on Selected
  Areas in Communications}, vol.~21, no.~5, pp. 793--801, 2003.

\bibitem{HeathIgnacio}
I.~Santamaria, O.~Gonazales, R.~W. Heath, Jr., and S.~W. Peters, ``Maximum
  sum-rate interference alignment algorithms for mimo channels,'' \emph{to
  appear in Proc. of IEEE Global Telecommunications Conference, Miami, Fl},
  Dec. 2010.

\bibitem{PetHea:Cooperative-algorithms-for-the-MIMO-Interference:09}
S.~W. Peters and R.~W. Heath, Jr., ``Cooperative algorithms for {MIMO}
  interference channels,'' \emph{CoRR}, vol. abs/1002.0424, 2010.

\bibitem{Rose_Greedy}
C.~Rose, S.~Ulukus, and R.~Yates, ``Wireless systems and interference
  avoidance,'' \emph{IEEE Trans. Wireless Commun.}, vol.~1, July 2002.

\bibitem{PaulrajMIMO}
A.~Paulraj, R.~Nabar, and D.~Gore, \emph{Intro. to Space-Time Wireless
  Communications}.\hskip 1em plus 0.5em minus 0.4em\relax New York, NY, USA:
  Cambridge University Press, 2008.

\bibitem{YeBlum}
S.~Ye and R.~Blum, ``Optimized signaling for {MIMO} interference systems with
  feedback,'' \emph{IEEE Trans. Signal Process.}, vol.~51, no.~11, pp.
  2839--2848, Nov 2003.

\bibitem{LabVIEW}
G.~Johnson and R.~Jennings, \emph{{LabVIEW graphical programming}}.\hskip 1em
  plus 0.5em minus 0.4em\relax McGraw-Hill Professional, 2006.

\bibitem{barhumi2003optimal}
I.~Barhumi, G.~Leus, and M.~Moonen, ``{Optimal training design for {MIMO OFDM}
  systems in mobile wireless channels},'' \emph{IEEE Trans. Signal Process.},
  vol.~51, no.~6, pp. 1615--1624, 2003.

\bibitem{synch}
E.~Zhou, X.~Zhang, H.~Zhao, and W.~Wang, ``Synchronization algorithms for {MIMO
  OFDM} systems,'' \emph{Proc. of IEEE Wireless Communications and Networking
  Conference, New Orleans, LA}, vol.~1, pp. 18--22 Vol. 1, March 2005.

\bibitem{1045}
\BIBentryALTinterwordspacing
{National Instruments}, ``{NI PXI-1045 Data Sheet},'' 2009. [Online].
  Available: \url{http://www.ni.com/pdf/products/us/pxi1045.pdf}
\BIBentrySTDinterwordspacing

\bibitem{5670}
\BIBentryALTinterwordspacing
------, ``{NI PXI-5670 Data Sheet},'' 2009. [Online]. Available:
  \url{http://www.ni.com/pdf/products/us/5670\_datasheet.pdf}
\BIBentrySTDinterwordspacing

\bibitem{5660}
\BIBentryALTinterwordspacing
------, ``{NI PXI-5660 Data Sheet},'' 2009. [Online]. Available:
  \url{http://www.ni.com/pdf/products/us/4mi469-471.pdf}
\BIBentrySTDinterwordspacing

\bibitem{RapidProto}
A.~Gupta, A.~Forenza, and R.~W. Heath, Jr., ``Rapid {MIMO-OFDM} software
  defined radio system prototyping,'' \emph{Proc. of IEEE Workshop on Signal
  Processing Systems}, October 2004.

\bibitem{6653}
\BIBentryALTinterwordspacing
{National Instruments}, ``{NI PXI-6653 Data Sheet}.'' [Online]. Available:
  \url{http://www.ni.com/pdf/products/us/pxi665x\_pxie6672\_datasheet.pdf}
\BIBentrySTDinterwordspacing

\bibitem{wallace2003experimental}
J.~Wallace, M.~Jensen, A.~Swindlehurst, and B.~Jeffs, ``{Experimental
  characterization of the {MIMO} wireless channel: Data acquisition and
  analysis},'' \emph{IEEE Trans. Wireless Commun.}, vol.~2, no.~2, pp.
  335--343, 2003.

\bibitem{czink2100can}
N.~Czink, B.~Bandemer, G.~Vilar, L.~Jalloul, and A.~Paulraj, ``{Can multi-user
  MIMO measurements be done using a single channel sounder?}'' \emph{COST 2100,
  TD (08), Lille, France, October 2008}, vol. 621.

\bibitem{golub1996matrix}
G.~Golub and C.~Van~Loan, \emph{{Matrix computations}}.\hskip 1em plus 0.5em
  minus 0.4em\relax Johns Hopkins Univ Pr, 1996.

\bibitem{edelman}
A.~Edelman, T.~A. Arias, and S.~T. Smith, ``The geometry of algorithms with
  orthogonality constraints,'' \emph{SIAM Journal on Matrix Analysis and
  Applications}, vol.~20, no.~2, pp. 303--353, 1998.

\bibitem{balanisantenna}
C.~A. Balanis, \emph{Antenna Theory: Analysis and Design}.\hskip 1em plus 0.5em
  minus 0.4em\relax New York, NY: Wiley, 1997.

\end{thebibliography}

\newpage

\begin{IEEEbiography}[{\includegraphics[width=1in,height=1.25in,clip,keepaspectratio]{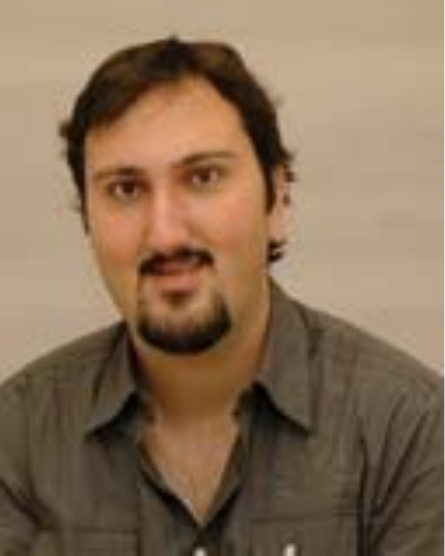}}]{Omar El Ayach}
(S'08) received his B.E. degree in computer and communications engineering from the American University of Beirut, Lebanon, in 2008. He completed his M.S. in electrical engineering degree at The University of Texas at Austin in 2010. 

He is now a Ph.D. student in the Wireless Networking and Communication Group (WNCG) at The University of Texas at Austin under the supervision of Prof. Robert W. Heath, Jr. in the Wireless Systems and Innovation Lab (WSIL) group. His research interests are in the broad area of MIMO signal processing and information theory, in particular interference management techniques in wireless systems and network sciences.
\end{IEEEbiography}

\vspace{-2.4in}

\begin{IEEEbiography}[{\includegraphics[width=1in,height=1.25in,clip,keepaspectratio]{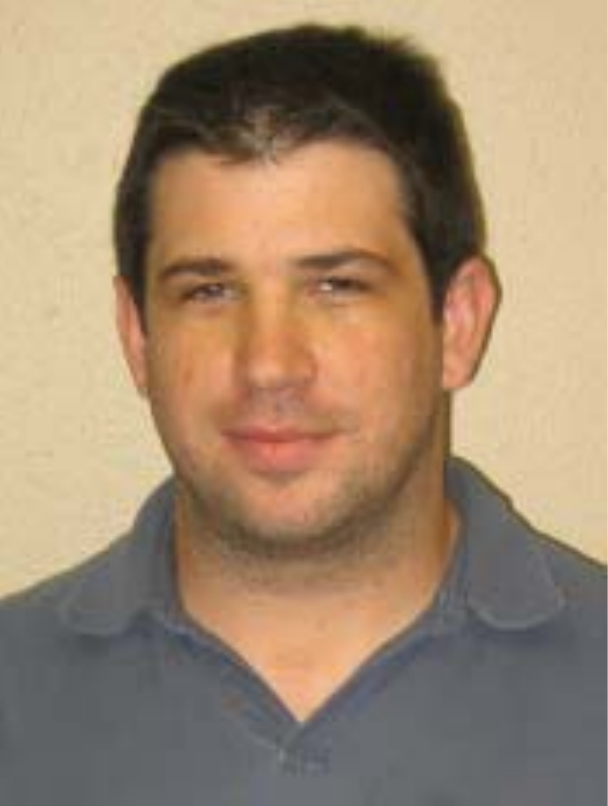}}]{Steven W. Peters}
 is a Ph.D. student at the University of Texas at Austin and
Co-Founder and CEO of Kuma Signals, LLC.
He received B.S.~degrees in electrical engineering and computer engineering
from the Illinois Institute of Technology in 2005 and an M.S.E.~degree
in electrical engineering from the University
of Texas at Austin in 2007. From 2005--2007 he was a research
assistant at the Applied Research Laboratories, where he worked
on physical layer and coding design for ground-wave high frequency
transhorizon communication systems. He has served as a consultant
to several companies working on wireless system design and standards.
His research interests include interference mitigation techniques in
wireless networks, cooperative communication, and MIMO.
\end{IEEEbiography}

\vspace{-2.4in}

\begin{IEEEbiography}[{\includegraphics[width=1in,height=1.25in,clip,keepaspectratio]{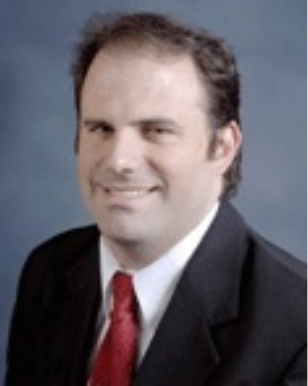}}]{Robert W. Heath, Jr.}
(S'96-M'01-SM'06) received the B.S. and M.S. degrees from the University of Virginia, Charlottesville, in 1996 and 1997, respectively, and the Ph.D. degree from Stanford University, Stanford, CA, in 2002, all in electrical engineering.

From 1998 to 2001, he was a Senior Member of the Technical Staff then a Senior Consultant with IospanWireless Inc., San Jose, CA, where he worked on the design and implementation of the physical and link layers of the first commercial MIMO-OFDM
communication system. In 2003, he founded MIMO Wireless Inc., a consulting company dedicated to the advancement of MIMO technology. Since January 2002, he has been with the Department of Electrical and Computer Engineering, The University of Texas at Austin, where he is currently an Associate Professor and Associate Director of the Wireless Networking and Communications Group. His research interests include several aspects of MIMO communication: limited feedback techniques, multihop networking, multiuser MIMO, antenna design, and scheduling algorithms, as well as 60-GHz communication techniques and multimedia signal processing. 

Prof. Heath has been an Editor for the IEEE TRANSACTIONS ON COMMUNICATIONS and an Associate Editor for the IEEE TRANSACTIONS ON VEHICULAR TECHNOLOGY. He is a member of the Signal Processing for Communications Technical Committee in the IEEE Signal Processing Society and is the Vice Chair of the IEEE COMSOC Communications Technical Theory Workshop, is a general organizer for the 2009 CAMSAP Conference, and was a Technical Co-Chair for the 2010 IEEE International Symposium on Information Theory. He is the recipient of the David and Doris Lybarger Endowed Faculty Fellowship in Engineering. He is a licensed Amateur Radio Operator and is a registered Professional Engineer in Texas.
\end{IEEEbiography}

\end{document}